\documentclass[
10pt,
twocolumn,
superscriptaddress,
groupedaddress,
 amsmath,amssymb,
 aps,
pra,
]{revtex4-2}

\usepackage{cancel, comment}

\usepackage{amsmath}
\usepackage{mathtools}
\usepackage{amsfonts}
\usepackage{physics}
\usepackage{dsfont}
\usepackage{bm}
\usepackage{stackengine} % for stacking text above and below
\usepackage{enumerate}
\usepackage[colorlinks=true,linkcolor=blue, citecolor=blue, urlcolor=blue, bookmarks]{hyperref}

\usepackage{arydshln}
\usepackage{graphicx}

\usepackage[caption=false]{subfig}
\usepackage{array}
\usepackage{arydshln} % for dashed line in array
\usepackage{graphicx}% Include figure files
\usepackage{tabularx} % for tables with adjustable column width
\usepackage{float}
\usepackage[table, svgnames]{xcolor}
\usepackage{adjustbox}
\usepackage{placeins}

\usepackage{tikz}
\usetikzlibrary{arrows,backgrounds,positioning,fit,matrix,angles,quotes}
\usepackage{pgfplots}
\pgfplotsset{compat=1.18}
\usetikzlibrary{external}

\DeclareMathOperator{\sign}{sgn}
\DeclareMathOperator*{\oprod}{\bigotimes} % tensor product
\newcommand\numberthis{\addtocounter{equation}{1}\tag{\theequation}} % for numbering final line of align* environment
\newcolumntype{L}[1]{>{\raggedright\let\newline\\\arraybackslash}p{#1}}
\newcolumntype{C}[1]{>{\centering\let\newline\\\arraybackslash}m{#1}}
\newcolumntype{R}[1]{>{\raggedleft\let\newline\\\arraybackslash}m{#1}}
\usepackage{arydshln}

\newcommand{\be}{\begin{equation}}
\newcommand{\ee}{\end{equation}}
\newcommand{\myvec}[1]{\boldsymbol{#1}}

\newcommand{\statetwo}[1]{
\hspace{2pt}
\begin{tikzpicture}[x=7pt,y=7pt,baseline=.1ex]
    \node (0) at (0,0) {};
    \node (1) at (1,0) {};
    \node (2) at (0,1) {};
    \node (3) at (1,1) {};
    \foreach \i in {1,...,3}{
        \draw (\i) circle(2pt);};
    \foreach \i in {#1}{
        \fill[black] (\i) circle(2.2pt);};
    \draw (1.5,-0.5) -- (2, 0.5);
    \draw (1.5,1.5) -- (2,0.5);
    \draw (-0.7,-0.5) -- (-0.7,1.5);
\end{tikzpicture}
\hspace{2pt}}

\newcommand{\stateone}[1]{
\hspace{2pt}
\begin{tikzpicture}[x=7pt,y=7pt,baseline=-.6ex]
    \node (0) at (0,0) {};
    \node (1) at (1,0) {};
    \node (2) at (2,0) {};
    \node (3) at (3,0) {};
    \foreach \i in {1,...,3}{
        \draw (\i) circle(2pt);};
    \foreach \i in {#1}{
        \fill[black] (\i) circle(2.2pt);};
    \draw (3.5,-0.5) -- (3.8, 0);
    \draw (3.5,0.5) -- (3.8, 0);
    \draw (-0.7,-0.5) -- (-0.7,0.5);
\end{tikzpicture}
\hspace{2pt}}

\begin{document}

\title{Quench dynamics in lattices above one dimension: the free fermionic case}

\author{Molly Gibbins}
\author{Arash Jafarizadeh}
\author{Adam Gammon-Smith}
\author{Bruno Bertini}

\affiliation{School of Physics and Astronomy, University of Nottingham, Nottingham, NG7 2RD, UK}
\affiliation{Centre for the Mathematics and Theoretical Physics of Quantum Non-Equilibrium Systems, University of Nottingham, Nottingham, NG7 2RD, UK}

\date{\today}

\begin{abstract}
We begin a systematic investigation of quench dynamics in higher-dimensional lattice systems considering the case of non-interacting fermions with conserved particle number. We prepare the system in a translational-invariant non-equilibrium initial state --- the simplest example being a classical configuration with fermions at fixed positions on the lattice --- and let it to evolve in time. We characterise the system's dynamics by measuring the entanglement between a finite connected region and its complement. We observe the transmutation of entanglement entropy into thermodynamic entropy and investigate how this process depends on the shape and orientation of the region with respect to the underlying lattice. Interestingly, we find that irregular regions display a distinctive multi-slope entanglement growth, while the dependence on the orientation angle is generically fairly weak. This is particularly true for regions with a large (discrete) rotational symmetry group. The main tool of our analysis is the celebrated quasiparticle picture of Calabrese and Cardy, which we generalise to describe the case at hand. Specifically, we show that for generic initial configurations (even when restricting to classical ones) one has to allow for the production of multiplets involving ${n>2}$ quasiparticles and carrying non-diagonal correlations. We obtain quantitatively accurate predictions --- tested against exact numerics --- and propose an efficient Monte Carlo-based scheme to evaluate them for arbitrary connected regions of generic higher dimensional lattices.  
\end{abstract}

\maketitle

\section{Introduction}

Finding an efficient description of quantum matter out of equilibrium is every bit as important and timely as it is difficult. Despite the first efforts to attack this problem dating back to the work of John von Neumann in the late 1920s~\cite{von2010proof}, almost one century later we are still lacking general tools to characterise quantum many-body dynamics in an effective and systematic fashion. 

Of course, the fact that this endeavour is difficult does not mean that no progress has been achieved. In the ninety four years following von Neumann's work, and especially during the last two decades, a remarkable effort has been directed to this problem and significant results have been obtained~\cite{calabrese2016introduction, calabrese2016introduction, vidmar2016generalized, essler2016quench, doyon2020lecture, bastianello2022introduction, alba2021generalized, PolkovnikovReview, gogolin2016equilibration, rigol2008thermalization, serbyn2021quantum, calabrese2016, bertini2021finitetemperature}. In particular, the case of one-dimensional systems turned out to be the one providing the most important advances. In this case it is possible to use powerful mathematical structures such as integrability~\cite{essler2016quench, doyon2020lecture, bastianello2022introduction, alba2021generalized, klobas2021exact, klobas2021entanglement, klobas2021exactrelaxation}, conformal invariance~\cite{quasi2005, calabrese2009entanglement, calabrese2016}, dual-unitarity~\cite{bertini2019exact, bertini2019entanglement, piroli2020exact}, and random circuit averaging~\cite{nahum2017quantum, nahum2018operator, vonKeyserlingk2018operator, zhou2019emergent, fisher2022random} to find analytical descriptions, while, at the same time, powerful numerical methods based on matrix product states~\cite{schollwoeck2011the, daley2004time, white2004real, vidal2003efficient, vidal2004efficient} are able to follow the evolution of generic many-body systems, at least for short times.

These remarkable tools allowed us to understand several consequential concepts pertaining to the dynamics and eventual relaxation of quantum matter out of equilibrium. One key realisation has been that the equilibration process in quantum many-body systems works \emph{locally in space}, i.e., local subsystems are eventually described by time-independent statistical ensembles even though the dynamics of the full system conserves probabilities~\cite{essler2016quench}. Another breakthrough has been to identify the quantum entanglement between a local subsystem and the rest as the ``universal observable'' able to characterise the full relaxation process in an elegant and basis-independent manner~\cite{amico2008entanglement, calabrese2009entanglement, laflorencie2016quantum, calabrese2018entanglement}. In essence, one can describe relaxation as the process of turning entanglement entropy into thermodynamic entropy~\cite{calabrese2018entanglement, santos2011entropy, gurarie2013global}. This process displays an astonishing universality across a huge spectrum of different locally interacting systems which has been explained as the result of a duality between space and time~\cite{bertini2022growth} (see also Refs.~\cite{ bertini2022nonequilibrium, bertini2023dynamics}). 

Having now sharpened our theoretical tools it is natural to wonder whether we can move on from the one-dimensional setting and start exploring higher dimensional cases. In this work we initiate this venture by studying the entanglement dynamics in a ${d>1}$ dimensional lattice system of non-interacting fermions with conserved particle number (tight-binding model), which is driven out of equilibrium by means of a global quantum quench protocol. The main tool of our analysis is the quasiparticle picture of Ref.~\cite{quasi2005}, which is based on the assumption that, after the quench, quantum correlations are transported throughout the system by pairs of correlated quasiparticles created by the quench. Supplemented with a few bits of microscopic data~\cite{alba2017entanglement}, this picture gives asymptotically exact predictions for the entanglement dynamics of free~\cite{quasi2005,fagotti2008evolution,castroalvaredo2019entanglement} and interacting integrable~\cite{alba2017entanglement, alba2018entanglement, alba2019entanglement} theories, where quasiparticle excitations are infinitely stable. Our work parallels similar studies carried out in the context of continuum quantum field theory~\cite{liu2014entanglement, casini2016spread, cotler2016entanglement}; see also Refs.~\cite{maraga2015coarsening, chiocchetta2015scaling, chiocchetta2016scaling, lemonik2016bosons, chiocchetta2017crossovers} for other field-theory studies of higher dimensional quenches. 

We show that, to describe the dynamics from translational invariant initial configurations where the unit cell contains $|\myvec{\nu}| = \nu_1 \hdots \nu_d$ sites, the quasiparticle picture has to be generalised in the spirit of Refs.~\cite{pair2018, hydro2018, bastianello2018spreading}. Namely, one has to admit that, instead of pairs, the correlated quasiparticles form an $n$-plet~\cite{pair2018}. Moreover, one has to account for the fact that the quasiparticles generically show complicated, off-diagonal correlations that can be determined by computing the ``particle entanglement''~\cite{haque2009entanglement} for a given bipartition of the multiplet~\cite{hydro2018, bastianello2018spreading}. Proceeding in this way we obtain quantitatively accurate predictions, which show how the entanglement entropy of a region is transformed into thermodynamic entropy by the time evolution. Then we discuss how this process depends on shape and orientation of the region. We find that regions that are more irregular --- characterised by different length scales --- display a distinctive multi-slope entanglement growth. On the other hand, we see that the dependence on the orientation becomes increasingly weaker as we increase the discrete-rotation symmetry group of the region. We test our predictions against exact numerics, and propose an efficient Monte Carlo-based scheme to compute entanglement dynamics for arbitrary connected regions of generic $d$-dimensional lattices.

This paper is organised as follows. In Section \ref{sec:setup}, we introduce the precise setting considered in this work. In Section \ref{sec:qppicture} we introduce the quasiparticle description of entanglement growth and test its predictions against exact numerics for various $d=1$ and $d=2$ states. In Section \ref{sec:results} we use the quasiparticle description to investigate the entanglement growth in $d \geq 1$ and, in particular, how the latter depends on shape and orientation of the subsystem. Finally, in Section \ref{sec:conclusions} we present our conclusions and outlook. Additional technical details are included in the appendix.

\section{Setup}
\label{sec:setup}

In this paper we study a global quantum quench protocol in which a many-body quantum system is prepared in a non-equilibrium initial state $\ket{\Psi}$ and let to evolve according to its own unitary dynamics. In this section we describe the specific system/initial state considered and define the observable of interest. 

\subsection{Hamiltonian}

We consider a system of spinless fermions arranged, for convenience, on a square lattice in $d$ spatial dimensions and linear size $L$. The dynamics are generated by the following Hamiltonian

\begin{equation}
\label{hamiltonian}
    H = J \sum_{\langle \myvec{n},\myvec{m} \rangle} \left( c^\dag_{\myvec{n}}  c_{\myvec{m}}^{\vphantom{\dag}} + {\rm h.c.} \right),
\end{equation}
where $J$ is the coupling strength, $\myvec{n}\in a \mathbb Z_L^d$ denotes a point on the $d$ dimensional lattice with spacing $a$, $\langle \myvec{n},\myvec{m} \rangle$ indicates that the sum is restricted to nearest neighbours, and finally  $c_{\myvec{n}}$ denote canonical fermionic operators. In the following, the lattice spacing $a$ will be set to $1$ unless explicitly stated.

The Hamiltonian \eqref{hamiltonian} is invariant under one-site translations and is diagonalised by Fourier transform 
\begin{align}
\label{spectrum}
&H = \sum_{\myvec{k}} \epsilon(\myvec{k}) \tilde c^\dag_{\myvec{k}} \tilde c_{\myvec{k}}^{\vphantom{\dag}}, & & \epsilon(\myvec{k}) = 2J\sum_{i=1}^d \cos(k_i)\,,
\end{align}
where $\{\tilde c^\dag_{\myvec{k}},\tilde c_{\myvec{k}}^{\vphantom{\dag}}\}$ are the Fourier transformed fermions, 
\be
\tilde{c}_{\myvec k}  = \frac{1}{L^{d/2}} \sum_{\myvec n \in  \mathbb Z^d_{L} } e^{i  {\myvec k}\cdot {\myvec n}} c_{\myvec n}\,,
\label{eq:modes}
\ee
$\myvec{k}\in (2\pi/L)\mathbb Z_L^d$ is a quasi-momentum in the $d$-dimensional Brillouin zone and $k_i$ denotes its $i$-th component. 

\subsection{Initial State}
\label{initial state section}

We focus on initial states that are Gaussian, low-entangled, invariant under $\nu$-site translations, and with fixed particle number. Namely we consider states of the form
\begin{equation}
\label{eq:initgeneral}
    \ket{\psi_{\myvec{\nu}}} = \oprod_{\myvec{j}\in \mathbb Z_{\myvec L/\myvec \nu}} \ket{\psi_{\myvec{\nu},\myvec{j}}},
\end{equation}
where $\ket*{\psi_{\myvec{\nu},\myvec{j}}}$ is written in terms of fermionic operators within the $\myvec{j}$-th unit cell. In the following, $\myvec{L}=(L, \ldots,L)$ and is a $d$-dimensional vector, and the operations among $d$ dimensional vectors are always intended element-wise $\myvec{L}/\myvec{\nu}=(L/\nu_1, \ldots,L/\nu_d)$, and, for a given vector $\myvec n\in \mathbb N^d$ we set
\begin{equation}
\mathbb Z_{\myvec n}\equiv \mathbb Z_{n_1}\times\cdots\times \mathbb Z_{n_d}.
\end{equation}

This paper will devote particular attention to the subset of these initial states, denoted by $\ket{\psi^{\rm c}_{\myvec{\nu}}}$, where the fermions are at fixed initial positions, i.e., they can be thought of as classical configurations. For these states we have 

\begin{equation}
\label{init}
    \ket*{\psi^{\rm c}_{\myvec{\nu},\myvec{j}}} = c^\dag_{\myvec{a}_1+ \myvec{\nu} \myvec{j}} \hdots c^\dag_{\myvec{a}_N+\myvec{\nu} \myvec{j}} \ket{0}, \quad \quad \myvec{a}_i \in \mathbb Z_{\myvec{\nu}}. 
\end{equation}

\begin{figure}[H]
\centering
    $(a)$
    \begin{minipage}[c]{1cm}
        \scalebox{.8}{\includegraphics{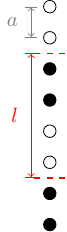}}
    \end{minipage}%
    \hspace{20pt}
    $(b)$
    \begin{minipage}[c]{3cm}
        \centering
        \scalebox{.8}{\includegraphics{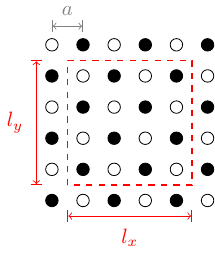}}
        \end{minipage}%
    \caption{Examples of initial states defined by Eqs. \eqref{eq:initgeneral} and \eqref{init}. On the accompanying diagrams, both the lattice spacing $a$, and an arbitrary subsystem of dimension $\myvec{l}$, are indicated for later reference. a) the $d=1$ state in Eq.~\eqref{eq:psi4}; b) The $d=2$ state in Eq.~\eqref{eq:psi22}.}
    \label{illustrations}
\end{figure}

Two concrete examples of these states, one in $d=1$ and one in $d=2$, are 
\begin{align}
        &\ket{\psi^{\rm c}_{4}} = \oprod_{{j}=0}^{{L}/{\nu}-1} c^\dag_{4 j} c^\dag_{1+4 j} \ket{0} \equiv \stateone{0,1}^{\otimes L/4}, \label{eq:psi4}\\
        &\ket{\psi^{\rm c}_{2,2}} = \oprod_{\myvec{j}=0}^{\myvec{L}/\myvec{\nu}-1} c^\dag_{(2,2)\cdot\myvec{j}} c^\dag_{(1,1)+(2,2)\cdot\myvec{j}} \ket{0} \equiv \statetwo{0,3}^{\otimes L^2/4}.  \label{eq:psi22}
\end{align}
A diagrammatic representation of these classical configurations is provided in Fig.~\ref{illustrations}. 

\subsection{Observable of Interest}
\label{sec:observable}

We characterise the evolution of the system by studying the dynamics of quantum entanglement between a chosen subsystem $A$ and the rest of the system $\bar{A}$. Since the state of the entire system is pure, the entanglement is conveniently measured by computing the \emph{entanglement entropy}, i.e., the von Neumann entropy of the reduced density matrix ${\rho}_A$ for the subsystem $A$~\cite{amico2008entanglement}. Namely, we consider 
\begin{equation}
\label{neumann}
    S_A(t) \equiv S({\rho}_A) = - \tr({\rho}_A \ln {\rho}_A).
\end{equation}
In a free-fermionic system evolving from a Gaussian state all correlations are encoded in the fermionic two-point functions. In particular, in our case the entanglement entropy is expressed as~\cite{review2009}
 \begin{equation}
    \label{contributions}
    S (\hat{\rho}_A) = - \tr\left[C_A\ln C_A \right] -\tr\left[(1-C_A)\ln(1-C_A)\right]\!,
\end{equation}
where $C_A$ is the correlation matrix of the subsystem $A$. The latter is obtained from the full correlation matrix 
\be
C_{\myvec{n},\myvec{m}} = \bra{\psi(t)}c^\dag_{\myvec{n}} c^{\phantom{\dag}}_{\myvec{m}} \ket{\psi(t)},
\ee
by eliminating rows and columns indexing particles in $\bar{A}$. As an example, in Appendix~\ref{app:numerics} we report the explicit form of $C_{\myvec{n},\myvec{m}}$ for the states \eqref{init}.

In a translational invariant, non-interacting system evolving from a Gaussian state the correlation matrix at time $t$ can be computed directly in the thermodynamic limit ($\bar A\to \infty$) with an amount of resources scaling polynomially with the number of sites of the subsystem $A$. Therefore, Eq.~\eqref{contributions} provides an efficient tool to characterise the entanglement dynamics. However, it does not provide direct insight into the relaxation process. To achieve the latter, in the next subsection we present a simple emergent description of the dynamics based on the propagation of stable quasiparticles~\cite{quasi2005}. We will show that, upon supplementing it with a small set of microscopic data, this quasiparticle picture provides an exact asymptotic description.

\section{Quasiparticle Picture}
\label{sec:qppicture}

The quasiparticle picture is based on the observation that non-interacting systems (but also interacting integrable ones~\cite{korepin1997quantum, takahashi1999thermodynamics}) feature stable quasiparticle excitations --- in our case these are simply the momentum modes in Eq.~\eqref{spectrum}. Following Ref.~\cite{quasi2005} one can then imagine that at $t=0$ the quench produces a finite density of quasiparticle excitations, which, upon spreading through the system for ${t>0}$, drive the relaxation process and the growth of entanglement. To elevate this idea to a quantitative description one needs to characterise the motion of the quasiparticles and how they ``carry" quantum correlations through the system. This will be our task for the rest of this section. The final result is reported in Eqs.~\eqref{eq:SAQP} and \eqref{eq:multipletcorrelationmatrix}, while in Sec.~\ref{sec:analysisQP} we test it against exact numerical results and verify that its asymptotic value coincides with the thermodynamic entropy.  

The motion of quasiparticles can be characterised straightforwardly. Since the system under examination is non-interacting, over large scales quasiparticles move like free classical particles and their trajectory is fully specified by their velocities~\footnote{In the case of interacting integrable models the quasiparticles undergo non-trivial scattering. Their scattering, however, is always elastic and its sole effect is to renormalise the quasiparticle velocities~\cite{alba2017entanglement}}. In our case the latter can be directly obtained by computing the group velocity of the momentum modes, i.e., 
\be
{\myvec v}(\myvec k)=\nabla\epsilon(\myvec{k}) = -2J\big(\sin(k_1),\sin(k_2),\hdots,\sin(k_d)\big)
%\begin{pmatrix}
%-2\sin(k_1) \\
%-2\sin(k_2)\\
%\vdots \\
%-2\sin(k_d) \\
%\end{pmatrix}\,.
\ee
Understanding how these modes contribute to the growth of entanglement, however, is far less straightforward and requires further physical insight. 

The key assumption of the quasiparticle picture is that, while moving, the modes generate entanglement in position space but not in momentum space: that is, they merely propagate correlations already present in the initial state~\cite{quasi2005}. Specifically, one assumes that the modes created at the same position are correlated as specified by the initial state and, while moving far apart, they spread this correlation through the system. This means that the entanglement between two regions can be obtained by finding all the multiplets of correlated modes shared between the two regions and summing up their contributions to the entanglement. Therefore, to find a quantitative prediction, one has to determine these quantities~\cite{pair2018}.  

The task is particularly simple when the correlated modes come in pairs. Indeed, Ref.~\cite{alba2017entanglement} showed that, in this case, the relevant contribution can be inferred from the stationary value of the entanglement entropy. Crucially, however, in our higher dimensional setting the initial states generically create correlations among more than $2$ modes. This can be seen by expressing the states in terms of the fermionic operators in momentum space, 

\be
\label{eq:initfouriergen}
\ket{\psi_{\myvec{\nu}}} = \oprod_{\myvec{p}\in \frac{2\pi}{L}\mathbb Z_{\myvec L/\myvec \nu}} \ket*{\tilde{\psi}_{\myvec{\nu},\myvec{p}}}\,,
\ee

\begin{figure}[t]
\includegraphics[width=0.45\textwidth]{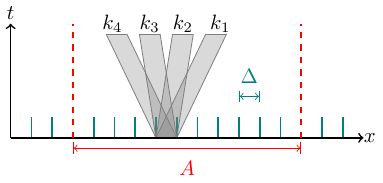}
\caption{Propagation of semiclassical cell modes for $d=1$ and $\nu=4$.}
\label{quasi evolution}
\end{figure}

For instance, considering the classical configurations \eqref{init} in Appendix~\ref{fourier transform} we find

\begin{equation}
\label{eq:initfourier}
  \ket*{\tilde {\psi}^{c}_{\myvec{\nu},\myvec{p}}} = \frac{1}{|\myvec \nu|^{N/2}} \prod_{p=1}^N \left( \sum_{\myvec{k}\in \frac{2\pi}{\myvec \nu}\mathbb Z_{\myvec{\nu}}} e^{-i \myvec{a}_p  \myvec{k}} \tilde c^{\dag}_{\myvec{p}+\myvec{k}} \right) \ket{0}, 
\end{equation}
where $|\myvec{\nu}| = \nu_1\dots \nu_d$ denotes the volume of the unit cell. As one can infer from this equation, these states generate correlations among $|\myvec \nu|$ modes. Requiring $|\myvec \nu|$ to be equal to 2 for $d\geq 2$ forces all $\nu_i$ but one to be equal to one. Namely, one is reduced to consider a state that is effectively one-dimensional. The same conclusion holds for all the states \eqref{eq:initgeneral}. Indeed, as shown in Appendix~\ref{gaussian}, they can all be expressed as in \eqref{eq:initfouriergen} where $ \ket*{\tilde{\psi}_{\myvec{\nu},\myvec{p}}}$ are written in terms of the fermionic operators with quasi momenta ${\myvec p}+\myvec k$ and $\myvec k \in ({2\pi}/{\myvec \nu})\mathbb Z_{\myvec{\nu}}$. 

Since, for these larger unit cells, one can no longer infer the entropy contributions from the stationary state as per  Ref.~\cite{alba2017entanglement}, we instead follow Ref.~\cite{hydro2018} (see also~\cite{bastianello2018spreading}) and reconstruct the evolution of the initial correlations using a semiclassical approach. This produces a closed form expression for the entanglement entropy valid at large scales and gives a natural way to compute the entanglement contributions.  

We begin by subdividing the system in hypercubic cells of linear size $\Delta$, which is much larger than the lattice spacing (one for us) but much smaller than the linear size of $A$, i.e.
\be
a \ll \Delta \ll |A|^{1/d}\,.
\label{eq:relevantlimit}
\ee
For convenience we take the hypercubic cells to be composed by an integer number of unit cells, i.e., $\myvec \Delta/\myvec \nu\in\mathbb N^d$. 

We then perform a partial Fourier transform to define the modes of the cell 
\be
\hat{c}_{\myvec x,\myvec k}  = \frac{1}{\Delta^{d/2}} \sum_{\myvec n \in  \mathbb Z^d_{\Delta} } e^{i  {\myvec k}\cdot {\myvec n}} c_{\Delta \myvec x+ \myvec n}, \qquad {\myvec k}\in \frac{2\pi}{\Delta} \mathbb Z^d_{\Delta}\,. 
\label{eq:cellmodes}
\ee

Intuitively, the idea is to define modes that are localised in both real and momentum space so that they can have well defined trajectories: see the schematic representation in Fig.~\ref{quasi evolution}. Writing the initial state in terms of the modes \eqref{eq:cellmodes} we have
\be
\ket{\psi_{\myvec{\nu}}} =  \oprod_{\myvec{x}\in \mathbb Z_{\myvec L/\myvec \Delta}} \oprod_{\myvec{p}\in \frac{2\pi}{\Delta}\mathbb Z_{\myvec \Delta/\myvec \nu}} \ket*{\hat{{\psi}}_{\myvec{\nu},(\myvec x, \myvec{p})}}\,,
\ee
where $\ket*{\hat{{\psi}}_{\myvec{\nu},(\myvec x, \myvec{p})}}$ coincides with $\ket*{\tilde{{\psi}}_{\myvec{\nu}, \myvec{p}}}$ in Eq.~\eqref{eq:initfouriergen} if one replaces ${\tilde c}_{\myvec p+\myvec k}$ with ${\hat c}_{\myvec x,\myvec p+\myvec k}$. For instance, for the classical configuration states \eqref{init} we have 
\begin{equation}
 \!\!\!\!\ket*{\hat{\psi}_{\myvec{\nu},(\myvec x, \myvec{p})}} \!=\! \frac{1}{|\myvec \nu|^{N/2}} \prod_{p=1}^N \left( \sum_{\myvec{k}\in \frac{2\pi}{\myvec \nu}\mathbb Z_{\myvec{\nu}}} e^{-i \myvec{a}_p  \myvec{k}} {\hat c}^\dag_{\myvec x,\myvec p+\myvec k} \right) \ket{0}.
\end{equation}
Assuming that the cell modes move classically in the limit \eqref{eq:relevantlimit}, the reduced density matrix of the subsystem $A$ can be computed by tracing out all the modes that are not in $A$ at time $t$, namely
\be
   \! \!\!\rho_A(t) \!\simeq\!\oprod_{\myvec{x}\in \mathbb Z_{{\myvec L}/{\myvec \Delta}}} \oprod_{\myvec{p}\in \frac{2\pi}{\Delta}\mathbb Z_{{\myvec \Delta}/{\myvec \nu}}}  \!\!\!\!\!\!\rho_{A}(\myvec{p},\myvec{x},t),
    \label{eq:semiclassicalrho}
\ee
where we introduced 
\begin{align}
&\rho_{A}(\myvec{p},\myvec{x},t)=\tr_{D_{A}(\myvec{p},\myvec{x},t)} \ketbra*{\hat{{\psi}}_{\myvec{\nu}, (\myvec{x}+\myvec{p})}},
\label{eq:bipartition}\\
&D_{A}(\myvec{p},\myvec{x},t) \!=\! \{(\myvec{x},\myvec{p}\!+\!\myvec k),\, \myvec k  \!\in\! \frac{2\pi}{\myvec \nu} \mathbb Z_{\myvec{\nu}}\!: \myvec{x} \!+\! v(\myvec{p}\!+\!\myvec k)t \notin A\}\,. \notag
\end{align}
In words, Eq.~\eqref{eq:semiclassicalrho} evaluates the reduced density matrix by tracing over the fermionic modes $(\myvec j, \myvec{p})$ that are out of the subsystem $A$. 

Plugging the expression \eqref{eq:semiclassicalrho} into the definition \eqref{neumann} we have 
\begin{equation}
    \!\!\!\!S_A (t) \simeq  \!\!\!\int_{\mathbb R^d} \!\!\!\!{\rm d}\myvec {x} \int_{\myvec{0}}^{\frac{\myvec{2\pi}}{\myvec{\nu}}} \!\!\!\frac{{\rm d} \myvec {p}}{\myvec{2\pi}} S(\rho_{A}(\myvec{p},\myvec{x},t)).
    \label{eq:SAQP}
\end{equation}
Eq.~\eqref{eq:SAQP} represents the desired quasi-particle expression for the entanglement entropy at time $t$. 

As we can see from Eq.~\eqref{eq:bipartition}, the contribution of the correlated multiplet represented by the momentum $\myvec{p}\in[\myvec{0}, {\myvec{2\pi}}/{\myvec{\nu}}]$ is found by computing the entanglement between the modes in $A$ and those out of it in the state $\ket*{\hat{{\psi}}_{\myvec{\nu}, (\myvec x, \myvec{p})}}$. This quantity is a measure of entanglement between modes, or particles, rather than between  regions of space and is referred to as \emph{particle entanglement}~\cite{haque2009entanglement}. In fact, since the state $\ket*{\hat{{\psi}}_{\myvec{\nu},(\myvec x, \myvec{p})}}$ is Gaussian, the entanglement can be computed using the fermionic correlation matrix as described in Sec.~\ref{sec:observable}. In particular, we define the $|\myvec{\nu}|\times|\myvec{\nu}|$ correlation matrix $\hat {C}(\myvec{p},\myvec{x})$ with matrix elements given by 
\begin{align}
\!\!\!\!\!\![\hat {C}(\myvec{p},\myvec{x})]_{\myvec{k},\myvec{k}'} \!\!=\!\! \expval*{ \hat{c}^\dag_{(\myvec x, \myvec{p}+\myvec k)} \hat c^{\phantom{\dag}}_{(\myvec x, \myvec{p}+\myvec k')}}{\hat {{\psi}}_{\myvec{\nu}, (\myvec x, \myvec{p})}}
\!,
\label{eq:correlmat}
\end{align}
where $\myvec k,  \myvec k'  \!\in\! (\myvec{2\pi}/{\myvec \nu}) \mathbb Z_{\myvec{\nu}}$. We then define the submatrix $\hat {C}_A(\myvec{p},\myvec{x},t)$ corresponding to the modes that are in $A$ at time $t$ by eliminating rows and columns corresponding to modes outside of $A$ at time $t$, i.e., $[\hat {C}(\myvec{p},\myvec{x})]_{\myvec{k},\myvec{k}'}$ such that ${(\myvec x, \myvec{p}+\myvec k),(\myvec x, \myvec{p}+\myvec k') \in D_{A}(\myvec{p},\myvec{x},t)}$. In terms of this submatrix we can finally write  
\begin{align}
   S(\rho_{A}(\myvec{p},\myvec{x},t)) &= -\tr\!\smash{[\hat {C}_A(\myvec{p},\myvec{x},t) \log\hat {C}_A(\myvec{p},\myvec{x},t)  ]}\label{eq:multipletcorrelationmatrix}\\
   &-\tr\!\smash{[(1\!-\!\hat {C}_A(\myvec{p},\myvec{x},t))\log\smash{(1\!-\!\hat {C}_A(\myvec{p},\myvec{x},t))}]}.\notag
\end{align}
It is worth emphasising that, since the dynamics of modes depends only on the unit cell size $\myvec \nu$, the exact structure of the initial state enters the entropy dynamics only through these entropy contributions. %The influence of this structure on the dynamics is explored further in Section~\ref{sec:results}. 

\begin{figure*}[t]
    \begin{minipage}[t]{\textwidth}
        \subfloat[]{%
            \begin{adjustbox}{width=0.5\textwidth}
                \includegraphics{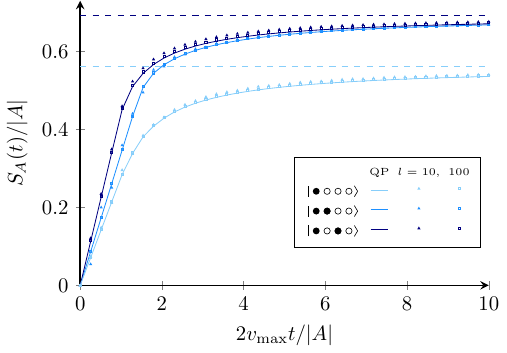}%
            \end{adjustbox}}
        \subfloat[]{%
            \begin{adjustbox}{width=0.5\textwidth}
                \includegraphics{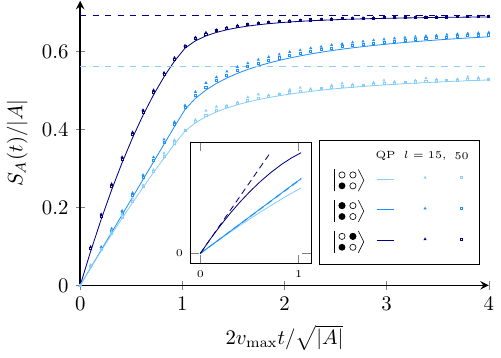}%
            \end{adjustbox}}
        \caption{Plots of the entropy over linear dimension $S_A(t)/|A|$ against rescaled time for (a) $d=1$ states with $\nu = 4$, and (b) $d=2$ states with $\nu_x = \nu_y = 4$ and a square subsystem. The states that are displayed reflect the number of `independent" states that do not map onto one other by single-site shifts of the unit cell, i.e. the number of states with unique entropy dynamics for a hypercubic subsystem. To compute the integral we employ an inverse FFT with (a) $10000$ subdivisions, and (b) $250 \times 250$ subdivisions. For each state, the quasiparticle solution (QP) is plotted against the finite size numeric solution with two different values of linear subsystem length $|A|=l$ in order to illustrate the rate of convergence. The dashed lines indicate the saturation values obtained from the stationary state solutions. The inset of (b) focuses on the initial regime, comparing each quasiparticle solution against a straight dashed line tangent to the solution at $t=0$.}
        \label{fig:QPtest}
    \end{minipage}%
\end{figure*}

\begin{figure}[b]
\centering
        \begin{adjustbox}{width=0.35\textwidth}
            \includegraphics{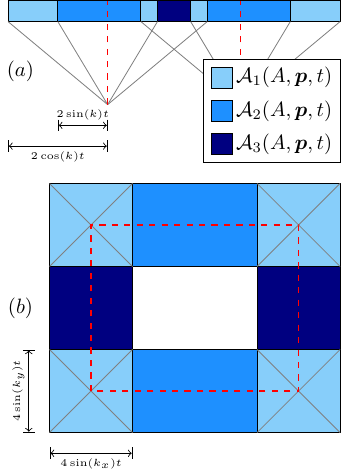}%
        \end{adjustbox}
        \caption{Diagram of the quasiparticle dynamics for a) a $d=1$ state with $\nu=4$, and b) a $d=2$ state with $\nu_x=\nu_y=2$ and rectangular subsystem, where $l_x>l_y$. The red dashed line marks the boundary of the subsystem and the three areas $\{{\cal A}_j(A, \bm{p},t)\}_{j=1,2,3}$ are depicted for a) $0<p<\pi/4$ and $l/2(\sin(p)+\cos(p))<t<l/4\sin(p)$, and b) $p_x=p_y$ and $t<l_x/4\sin(p_x)$. These areas are obtained by tracing the motion of quasiparticles as outlined in Section \ref{sec:qppicture}, and their explicit solutions are given in Appendix \ref{app:dynamics}.}
        \label{fig:areas}
\end{figure}

The integral in Eq.~\eqref{eq:SAQP} is conveniently evaluated by tracing the motion of quasiparticles. Namely, for fixed $(\myvec{p},t)$, one can trace the backward light cone of each mode $\myvec{p}+\myvec{k}$ being inside the subsystem at time $t$ as a displacement of the subsystem by $-v_{\myvec{p+k}}t$. The overlapping regions of these light cones indicate the origin of multiplets that have multiple modes inside the subsystem at time $t$, i.e.\ those corresponding to ${(\myvec x, \myvec{p}+\myvec k) \notin D_{A}(\myvec{p},\myvec{x},t)}$, see Fig.~\ref{fig:areas} for explicit examples. Proceeding in this way we can single out all possible splittings of correlated cell modes contributing to the entanglement --- we call them \emph{cell mode bipartitions} --- and the spatial regions where they are produced. The entanglement contribution of each splitting is then evaluated via Eq.~\eqref{eq:multipletcorrelationmatrix}. In summary, we can rewrite Eq.~\eqref{eq:SAQP} as follows  
\begin{align}
    S_A (t) &= \int_{0}^{\frac{\myvec{2\pi}}{\myvec{\nu}}} \!\!\!\frac{{\rm d} \myvec {p}}{\myvec{2\pi}} \sum_{a\in \cal B}s_a(\bm p) {\cal A}_a(A, \bm{p},t), 
\label{eq:xintegratedqp}
\end{align}
where $\mathcal{B}$ is the set of all \emph{cell mode bipartitions}, $s_a(\bm p)$ is the entanglement contribution of the bipartition $a\in \mathcal{B}$, and ${\cal A}_a(A, \bm{p},t)\geq0$ is the area of the spatial region producing multiplets contributing to the bipartition $a\in \mathcal{B}$ at time $t$ and for momentum $\bm p$. If a specific bipartition $a\in \mathcal{B}$ does not appear for a given choice of $(\myvec{p},t)$ we have ${\cal A}_a(A, \bm{p},t)=0$. 

To gain a clearer understanding of how Eq. \eqref{eq:xintegratedqp} is put into practice, we refer the reader to Section \ref{sec:resultsbasic}, which considers the simplest irreducible form of these areas in $d=2$ arising from a hypercubic lattice and subsystem and the choice $\nu_x, \nu_y = 2$. In this section, helpful illustrations are given for the areas ${\cal A}_a(A, \bm{p},t)$ in both this and the analogous $d=1$ case, as well as the explicit solution to one of these areas; the complete set of solutions is deferred to Appendix \ref{app:dynamics}.

For simple shapes, the areas ${\cal A}_a(A, \bm{p},t)$ of the various regions can be determined analytically as a function of $(\myvec{p},t)$. Computing the corresponding entanglement contributions and integrating Eq.~\eqref{eq:xintegratedqp} numerically over $\myvec{p}$ one can determine the time evolution of $S_A (t)$: Section \ref{sec:resultsbasic} gives a practical example. Instead, for more general regions we evaluate both the areas ${\cal A}_a(A, \bm{p},t)$ and the integral over $\myvec{p}$ by means of a convenient Monte Carlo scheme that we detail in Appendix~\ref{app:MC}.

\subsection{Test of the Quasiparticle Formula}
\label{sec:analysisQP}

The quasiparticle formula (\ref{eq:SAQP}, \ref{eq:multipletcorrelationmatrix}) may be tested against exact numerical results for finite subsystems $A$, whereby the real space correlation matrix of the subsystem is obtained in the thermodynamic limit and used in Eq.~\eqref{contributions} --- a representative example of the comparison is provided in Fig.~\ref{fig:QPtest}. As shown in the figure, the finite size numerics approach the quasiparticle prediction in the scaling limit

\be
t, |A| \to \infty, \qquad t/|A| = {\rm fixed}. 
\label{eq:scaling}
\ee
This is in agreement with the expectation that the quasiparticle description becomes asymptotically exact in this limit.

Another important test of the quasiparticle solution concerns the long-time behaviour. Indeed, the entanglement entropy is known to approach  thermodynamic entropy as time increases~\cite{calabrese2018entanglement, santos2011entropy, gurarie2013global}. This means that the infinite time value of Eqs.~(\ref{eq:SAQP}, \ref{eq:multipletcorrelationmatrix}) should coincide with the thermodynamic entropy of the stationary state reached by the subsystem $A$. In our case the latter is given by 
\begin{equation}
    \!\!\frac{S_{\rm th}}{|A|} \simeq  - \!\!\!\int \!\!\frac{{\rm d} \myvec {p}}{\myvec{2\pi}} [n(\myvec{p})\log n(\myvec{p})\!+\!(1\!-\!n(\myvec{p}))\log(1\!-\!n(\myvec{p}))],
    \label{eq:SGGE}
\end{equation}
where the integral is over the full Brillouin zone and $n(\myvec{p})$ is the occupation number of the conserved momentum mode $\myvec{p}$, i.e., 
\be
n(\myvec{p}) = \expval*{ \tilde{c}^\dag_{\myvec{p}} \tilde c^{\phantom{\dag}}_{\myvec{p}}}{\psi}\,. 
\label{eq:occupationnumber}
\ee
Eq.~\eqref{eq:SGGE} is recovered by Eqs.~(\ref{eq:SAQP}, \ref{eq:multipletcorrelationmatrix}) by noting that for $t=\infty$ the only cell mode bipartitions contributing are those where a single mode of the multiplet is in the system and all the others are outside. This is because, as the modes have different velocities, those starting at the same position are infinitely far from each other at $t=\infty$ and only one of them can be in $A$. In this limit the matrix $\hat {C}_A(\myvec{p},\myvec{x},t)$ becomes $1\times 1$ and coincides with the occupation number of the only mode in the system. Namely, 
\be
\hat {C}_A(\myvec{p},\myvec{x},t) = \sum_{\myvec{k}\in \frac{2\pi}{\myvec \nu}\mathbb Z_{\myvec{\nu}}} \chi_A(\myvec{x} \!+\! v(\myvec{p}\!+\!\myvec k)t) n(\myvec{p}+\myvec{k}),
\ee
where $\chi_A(\myvec{x})$ is the characteristic function of $A$, i.e., ${\chi_A(\myvec{x}\in A)=1}$ and ${\chi_A(\myvec{x}\notin A)=0}$. Plugging back into (\ref{eq:SAQP}, \ref{eq:multipletcorrelationmatrix}) we then have 
\begin{align}
\!\!\!\!\frac{S_A (\infty)}{|A|} \simeq&  \! - \!\!\!\!\!  \sum_{{\myvec{k}\in \frac{2\pi}{\myvec \nu}\mathbb Z_{\myvec{\nu}}}}\!\!\int_{\myvec{0}}^{\frac{\myvec{2\pi}}{\myvec{\nu}}} \!\!\!\frac{{\rm d} \myvec {p}}{\myvec{2\pi}} [n(\myvec{p}\!+\!\myvec{k})\log n(\myvec{p}\!+\!\myvec{k})\!\notag\\
&\qquad\quad +\!(1\!-\!n(\myvec{p}\!+\!\myvec{k}))\log(1\!-\!n(\myvec{p}\!+\!\myvec{k}))]\\
=& \! - \!\!\! \int \!\!\!\frac{{\rm d} \myvec {p}}{\myvec{2\pi}} [n(\myvec{p})\log n(\myvec{p})\!+\!(1\!-\!n(\myvec{p}))\log(1\!-\!n(\myvec{p}))]\,,\notag
\end{align}
where in the second step we combined the $|\myvec \nu|$ integrals over reduced Brillouin zones into a single integral over the whole zone.

\section{Results}
\label{sec:results}

In this section we present the predictions for the growth of the entanglement entropy between a region $A$ and the rest of the system after a quench from simple translation-invariant states in $d=1$ and $d=2$. First we consider the case of $A$ being a simple hyper cubical region aligned with the lattice axes. Then, we investigate the effect of rotations with respect to the lattice. Finally, we study the entanglement growth for more general, irregular-shaped regions.

\subsection{$d=1$ and $d=2$ Square Subsystems}
\label{sec:resultsbasic}

We begin our discussion considering the case of $A$ being either a connected segment in $d=1$ or a rectangular region aligned with the lattice (with edges in the $x$ and $y$ directions) in $d=2$. To fix the ideas we consider cases producing four correlated modes:
\begin{itemize}
\item[(i)] $d=1$ and $\nu=4$;
\item[(ii)] $d=2$ and $\nu_x=\nu_y = 2$;
\end{itemize}
and look at the following classical configurations 
\be
\begin{aligned}
&\stateone{0},\stateone{0,1},\stateone{0,2},\\
&\statetwo{0}, \statetwo{0,3},  \statetwo{0,2}.
\end{aligned}
\label{eq:examplecc}
\ee
Note that the last two states of each row can be written with a smaller unit cell, $\nu=2$ and $(\nu_x,\nu_y)=(2,1)$ respectively. This means that their entanglement dynamics can also be described by pairs of quasiparticles, which is not the case for the other four.

\begin{figure*}[t]
    \begin{minipage}[t]{\textwidth}
        \subfloat[]{%
            \begin{adjustbox}{width=0.49\textwidth}
                \includegraphics{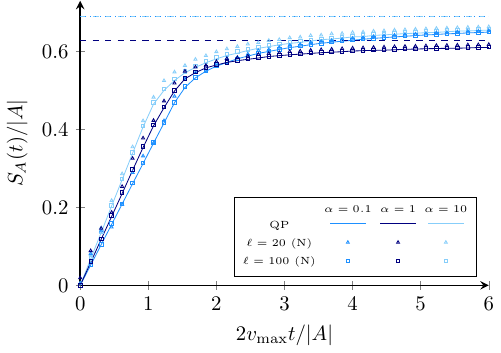}%
            \end{adjustbox}}
        \subfloat[]{%
            \begin{adjustbox}{width=0.49\textwidth}
                \includegraphics{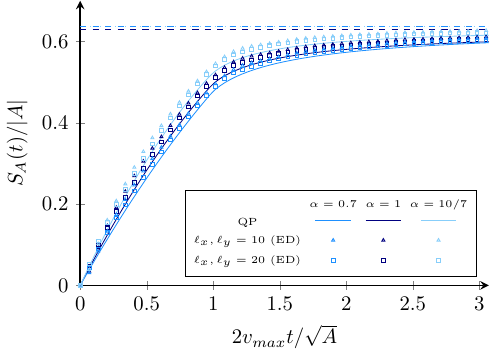}%
            \end{adjustbox}}
        \caption{Plots of the entropy over linear dimension $S_A(t)/|A|$ against rescaled time for (a) $d=1$ superposition initial states with $\nu = 4$
        such as $\ket{\varphi_4}\propto (\ket{\bullet\!\bullet\!\circ\circ}+\alpha\ket{\bullet\!\circ\!\bullet\circ})^{\otimes L/4}$, %\eqref{eq:phi-sup-1D}, 
        and (b) the $d=2$ superposition state  %$\statetwo{0,3}+\alpha\statetwo{1,2}$
        with $\nu_x = \nu_y = 2$ such as $\ket{\varphi_{2,2}}\propto (\ket{\bullet\!\circ\! //\! \bullet\!\circ}+\alpha\ket{\bullet\!\circ\! //\! \circ\!\bullet})^{\otimes L^2/4}$ %\eqref{eq:phi-sup-2D} 
        and a square subsystem. In (a), the integral is computed with inverse FFT with $10000$ subdivisions. In (b), exact diagonalization method for $L_x,L_y=80$ and $l_x,l_y=10,20$ was used. For each $\alpha$, the quasiparticle solution (QP) is plotted against the numeric solution to show the convergence rate. The QP solution was obtained by integrating the quasi-particle expression for the entanglement entropy at each time step.  The dashed lines indicate the saturation values for each state. The saturation value is identical for $\alpha$ and $\frac{1}{\alpha}$ in both the $d=1$ and $d=2$ cases.
        }\label{fig:QP Sup tests}
    \end{minipage}%    
\end{figure*}

In all these cases, the entanglement entropy can be efficiently computed by tracing the motion of the quasiparticles as described in Section \ref{sec:qppicture} (cf. Eq.~\eqref{eq:xintegratedqp}). An explicit example of this is shown in Fig.~\ref{fig:areas}. In particular, for all the states \eqref{eq:examplecc}, the relevant cell bipartitions produce three distinct entanglement contributions, $\{s_a(\bm{p})\}_{a=1,2,3}$. This effectively means that we have to specify only three areas $\{{\cal A}_a(A, \bm{p},t)\}_{a=1,2,3}$: see Figs.~\ref{fig:areas}(a) and ~\ref{fig:areas}(b) for illustrations of the three areas in $d=1$ and $d=2$ respectively, where in the latter we consider the more general case of $l_x \neq l_y$ to anticipate the section that follows. An example of the explicit form for one of these areas, ${\cal A}_2(A, \bm{p},t)$ from Fig.~\ref{fig:areas}(b), is given by
\be
\begin{aligned}
    {\cal A}_2(A, \bm{p},t) &=
    2(l_x-X)Y H(\min(\tau_x,\tau_y)-t) \\ &+ 2(l_x-X)l_y H(t-\tau_y)H(\tau_x-t)
\end{aligned}
\ee
where 
\begin{equation*}
    \tau_i \equiv l_i/4\sin(k_i), \quad X,Y \equiv 4\sin(k_i)t \quad \text{for} \quad i = x,y
\end{equation*}
The explicit form for all areas ${\cal A}_a(A, \bm{p},t)$ is reported in Appendix~\ref{app:dynamics}, while the entanglement contributions are reported in Tables~\ref{tbl:1d entropy contributions} and \ref{tbl:2d entropy contributions}. The resulting quasiparticle predictions are compared to the exact numerical solutions, in Fig.~\ref{fig:QPtest}. The left and right panels correspond respectively to $d=1$ and $d=2$.

\begin{table}[H]
\centering
    \begin{tabular}{ | C{1cm} | C{3cm} C{1.5cm} C{1.5cm} | }
        \hline
        & $\stateone{0}$ & $\stateone{0,1}$ & $\stateone{0,2}$ \\
        \hline
        $s_1$ & $2\ln 2 - 3\ln 3/4$ & $\ln 2$ & $\ln 2$ \\
        $s_2$ & $ \ln 2 $
        & $*$ & $2\ln 2$ \\
        $s_3$ & $\ln 2$ & $2\ln 2$ & $2\ln 2$ \\
        \hline
    \end{tabular}
\caption{The entropy contributions $\{s_j(\bm{p})\}_{j=1,2,3}$, defined by Eq. \eqref{eq:xintegratedqp} and Fig. \ref{fig:areas}a), for the three independent $\nu=4$ states of Eq. \eqref{eq:examplecc}. The explicit value of $*$ is $4\ln 2 - ({2+\sqrt{2}})/{2}\ln\left(({2+\sqrt{2}})/{4}\right) - ({2-\sqrt{2}})/{2}\ln\left(({2-\sqrt{2}})/{4}\right)$.}
\label{tbl:1d entropy contributions}
\end{table}

\begin{table}[H]
\centering
    \begin{tabular}{ | C{1cm} | C{3cm} C{1.5cm} C{1.5cm} | }
        \hline
        & $\statetwo{0}$ & $\statetwo{0,2}$ & $\statetwo{0,3}$ \\
        \hline
        $s_1$ & $2\ln 2 - 3\ln 3/4$ & $\ln 2$ & $\ln 2$ \\
        $s_2$ & $\ln 2$ & $0$ & $2\ln 2$ \\
        $s_3$ & $\ln 2$ & $2\ln 2$ & $2 \ln 2$ \\
        \hline
    \end{tabular}
\caption{The entropy contributions $\{s_j(\bm{p})\}_{j=1,2,3}$, defined by Eq. \eqref{eq:xintegratedqp} and Fig. \ref{fig:areas}b), for the three independent $\nu=4$ states of Eq. \eqref{eq:examplecc}.}
\label{tbl:2d entropy contributions}
\end{table}

Apart from the agreement between quasiparticle solution and exact numerics for increasing system sizes, which we have already stressed in Sec.~\ref{sec:analysisQP}, these figures show two significant features. First, we see that the $d=2$ case is clearly distinguished by the occurrence of a \textit{nonlinear} initial regime (see the inset of Fig.~\ref{fig:QPtest}). Indeed, using the explicit form of ${\cal A}_a(A, \bm{p},t)$ we see that Eq.~\eqref{eq:xintegratedqp} contains a quadratic term in time proportional to $(2s_1-s_2-s_3)$. From the entropy contributions presented in Table \ref{tbl:2d entropy contributions}, we see that the magnitude of this term is largest for $\statetwo{0,3}$, while it is zero for $\statetwo{0,2}$. This is consistent with the fact that the latter is effectively a one-dimensional setting. 

Another key takeaway from Fig.~\ref{fig:QPtest} is the difference between the entropy plots of states with equal occupation numbers \eqref{eq:occupationnumber} (the last two states in both lines of Eq.~\eqref{eq:examplecc} have $n(\myvec{p}) = {1}/{2}$). Examples include the slower initial growth of $\stateone{0,1}$ versus $\stateone{0,2}$ in Fig. \ref{fig:QPtest}(a), and of 
$\statetwo{0,2}$ versus $\statetwo{0,3}$ in Fig. \ref{fig:QPtest}(b), which share the same saturation value. One can offer a heuristic explanation of this slower growth: states of equal charge density but less uniform site occupation impose more constraints on the initial site hoppings; the plots indicate that these constraints last until correlated quasiparticles first span the full width of the subsystem. Importantly, since occupation numbers fully specify the expectation value of all conserved charges, our result shows that the expectation values of all conserved charges are not enough to determine the full-time entanglement dynamics even in the scaling regime \eqref{eq:scaling}.

A stark feature of Tables~\ref{tbl:1d entropy contributions} and \ref{tbl:2d entropy contributions} is that the classical configurations \eqref{eq:examplecc} turn out to give momentum independent entanglement contributions. To make sure that this property does not introduce any qualitative difference in the entanglement dynamics we also consider more general, non-classical states described by Eq. \ref{eq:initgeneral}, where $\ket*{\psi_{\myvec{\nu},\myvec{j}}}$ is now a superposition of states in the $|\myvec{\nu}|$-site unit cell that satisfies the Gaussianity condition given by Eq. \ref{eq:gaussianity}. The entanglement contributions for these superposition states are reported in Appendix~\ref{app:contributions} while Fig.~\ref{fig:QP Sup tests} reports some representative examples. Although in this case the entanglement contributions become momentum dependent (cf. App.~\ref{app:contributions}), we see that the plots are qualitatively similar to those in Fig.~\ref{fig:QPtest} and the agreement with the exact numerical solution is still excellent for large enough subsystems.

\begin{figure}[b]
\includegraphics{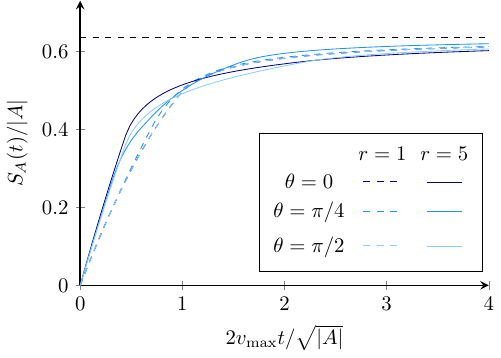}
\caption{Plot of the entropy over linear dimension $S_A(t)/|A|$ against rescaled time for the superposition initial state $(\ket{\bullet\!\circ\! //\! \bullet\!\circ}+\alpha\ket{\bullet\!\circ\! //\! \circ\!\bullet})^{\otimes L^2/4}$ %\eqref{eq:phi-sup-2D} 
with $\alpha=10/7$ and a rectangle-shaped subsystem with sides $\sqrt{r}$ and $\frac{1}{\sqrt{r}}$, for $r=1,5$ as angle of rotation $\theta$ is varied. The dashed line shows the saturation value $S_{A}(\infty)=0.63632$ for $\alpha=10/7$.}
\label{fig:rotatedrect}
\end{figure}

\begin{figure*}[t]
    \begin{minipage}[t]{\textwidth}
        \subfloat[]{%
            \begin{adjustbox}{width=0.33\textwidth}
                \includegraphics{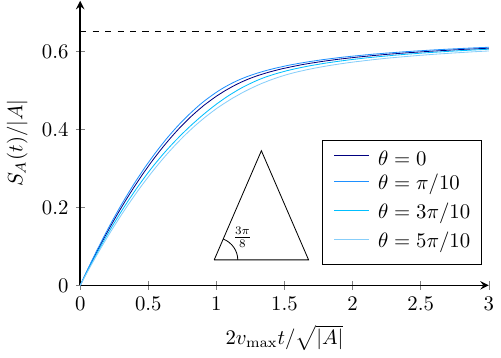}%
            \end{adjustbox}}
        \subfloat[]{%
            \begin{adjustbox}{width=0.33\textwidth}
                \includegraphics{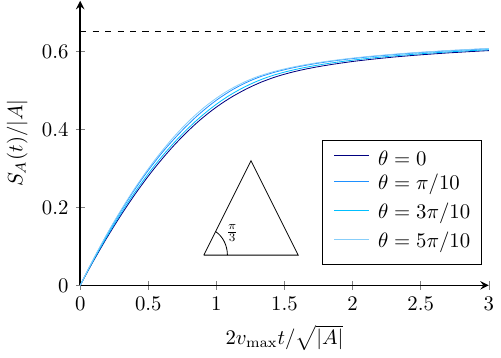}%
            \end{adjustbox}}
        \subfloat[]{%
            \begin{adjustbox}{width=0.33\textwidth}
                \includegraphics{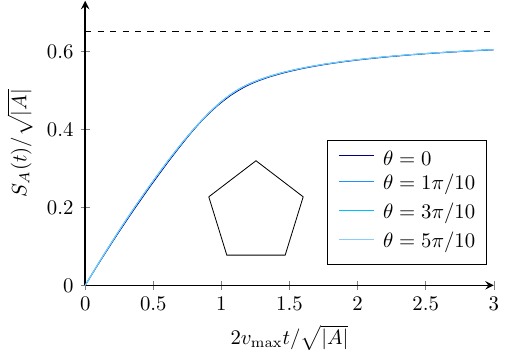}%
            \end{adjustbox}}
        \caption{Entropy density $S_A(t)$ against rescaled time for the superposition initial states $(\ket{\bullet\!\circ\! //\! \bullet\!\circ}+\alpha\ket{\bullet\!\circ\! //\! \circ\!\bullet})^{\otimes L^2/4}$ %Eq.~\eqref{eq:phi-sup-2D} 
        with $\alpha=0.5$ and different subsystems, as angle of rotation $\theta$ is varied. For each subsystem considered here, the shape of the region is demonstrated in the inset: (a) isosceles triangle; (b) equilateral triangle; (c) regular pentagon. The dashed line shows the saturation value given by the stationary state solution $S_{A}(\infty)=0.652301$ for $\alpha=0.5$.} 
        \label{fig:rotatepolygons}
    \end{minipage}%
\end{figure*}

\subsection{Rotations with respect to the lattice}
\label{sec:rotate}

Let us now consider the dependence of the entanglement growth on the orientation of $A$ with respect to the underlying lattice. We begin considering the simple case of a rectangular region in $d=2$ that is rotated by an angle $\theta$ with respect to the lattice.

In this case the explicit calculation of the areas in Eq.~\eqref{eq:xintegratedqp} becomes quite tedious and the quasiparticle prediction is more conveniently obtained integrating Eq.~\eqref{eq:SAQP} via the Monte Carlo scheme discussed in App.~\ref{app:MC}, which agrees with the explicit approach. The results for a representative initial state are reported in Fig.~\ref{fig:rotatedrect}. From Fig.~\ref{fig:rotatedrect}(a) we see that when the aspect ratio of the rectangle $r\gg 1$ the entanglement dynamics depends quite markedly on the orientation of the region, with the rectangle aligned with the lattice showing a slower relaxation. Interestingly, however, we see that the dependence on rotation angle decreases smoothly with the aspect ratio $r$ such that whenever the edges $l_x$ and $l_y$ coincide the dependence on the rotation angle almost disappears.

To exclude that this is not an artefact of the Monte Carlo integration routine we reproduced the result for the rotated square region using the quasiparticle tracing integration of Eq.~\eqref{eq:xintegratedqp} finding exact agreement. In this case the relevant areas depend on the rotation angle, see Fig.~\ref{appfig:areasrotate}, and their explicit expression is reported in Appendix~\ref{app:dynamics} while the entanglement contribution associated with each area is reported in Tabs.~\ref{tbl:1d entropy contributions} and ~\ref{tbl:2d entropy contributions} of the previous section.

\iffalse
\begin{figure}[H]
\centering
    \includegraphics{figures/rectangle_areas_rotate.pdf}%
    \caption{
    Diagram of the quasiparticle dynamics for a $\nu_x=\nu_y=2$ initial state and rotated rectangular subsystem with $l_x>l_y$. The red dashed line marks the boundary of the subsystem and the three areas ${\cal A}_j(A, \theta, \bm{p},t)$, $j=1,2,3$, are depicted for fixed $(\bm{p}, \theta, t)$. In this case, the structure of each of these shaded regions changes with angle of rotation $\theta$ such that the full solutions to ${\cal A}_j(A, \theta, \bm{p},t)$ $j=1,\ldots,3$ are piecewise functions in $\myvec{p}$. In particular, this diagram displays the structure for all $\myvec{p}$ such that $\sin(p_y)/\sin(p_x) \in [\theta,\pi/2-\theta]$.
    \label{fig:rotation}}
\end{figure}  
\fi

The behaviour in Fig.~\ref{fig:rotatedrect} can be expected as the square is ``more rotationally symmetric" than the rectangle. More precisely, it is left invariant by greater number of discrete rotations: its cyclic group is of order $4$ rather than $2$. To highlight how the order of the cyclic group of $A$ affects the orientation dependence, in Fig.~\ref{fig:rotatepolygons} we consider polygons with cyclic group of order $q=1,3,5$. We see that, as expected, the dependence on $\theta$ decreases with $q$. A surprising aspect, however, is how quickly it does so: the $\theta$ dependence is already negligible for $q=3$.

\begin{figure}[b]
    \includegraphics{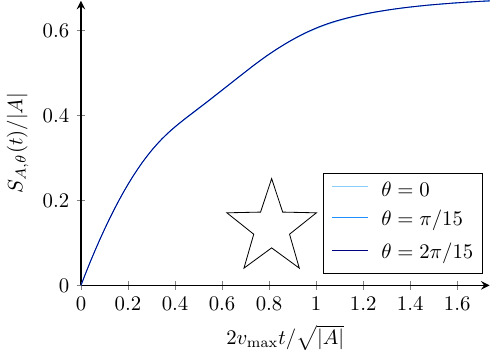}%
    \caption{Plot of the entropy density $S_A(t)$ against rescaled time $4t/|A|$ for the initial state $\ket{\bullet\circ//\circ\bullet}\,$ and a star-shaped subsystem, as angle of rotation $\theta$ is varied.} 
    \label{fig:star}
\end{figure}

\subsection{General Shapes}
\label{sec:generalshapes}

Finally, we use our quasiparticle approach to investigate the the entanglement growth of irregular, connected, regions characterised by different cross sections. Interestingly, we observe that such regions display an entanglement dynamics that is qualitatively different from that of regular ones. 

For instance,  in Fig.~\ref{fig:star} we show the entanglement dynamics of a region in the shape of five point star (see inset panel of Fig.~\ref{fig:star} for an illustration) for different orientation angles $\theta$ with respect to the underlying lattice. We see that, as expected, there is essentially no dependence on $\theta$, however, the entanglement evolution reported in the figure is quite peculiar: rather than the usual linear increase followed by saturation the entanglement shows a complicated multi-slope curve. 

This is due to the fact that each cross section of the figure corresponds to a non-analyticity of the quasiparticle prediction in time~\footnote{We remark that the presence of points of non-analyticity in the quasiparticle prediction is not in contradiction with the fact that, for any finite subsystem, the entanglement dynamics is smooth. Indeed, the quasiparticle prediction describes the asymptotic limit Eq.~\eqref{eq:scaling}.}. These special moments correspond to the points in time where the backward light cones associated to the fastest quasiparticles separate through a cross section - in other words, when a sudden gap appears between the light cones. For instance, for a rectangular region and the states in the second line of Eq.~\eqref{eq:examplecc}, the quasiparticle solution obtained integrating Eq.~\eqref{eq:xintegratedqp} with the ${\cal A}_a(A, \bm{p},t)$ in Appendix~\ref{app:dynamics} reads as 
\begin{align}
\frac{S_A(t)}{ l_x l_y} =&  4 f(\zeta \sqrt{r})f(\zeta/\sqrt{r}) s_1 +  f(\zeta \sqrt{r})(1-2 f(\zeta/\sqrt{r})) s_2 \notag\\
&+ (1-2f(\zeta \sqrt{r})) f(\zeta/\sqrt{r}) s_3, \label{eq:SA}
\end{align}
where we introduced $\zeta=2 v_{\rm max} t/\sqrt{l_x l_y}$, $r=l_x/l_y$, and 
\be
\!\!\!\! f(z) \!=\! 
\begin{cases}
\frac{1}{\pi} z & z \leq 1 \\
\frac{1}{2} - \frac{1}{\pi} \arcsin(\frac{1}{z}) + \frac{1}{\pi} (z - \sqrt{z^2 - 1}) & z>1 
\end{cases}\!.
\ee
This function has a non-analytic points (corresponding to a discontinuous second derivative) for $z=1$: this means that \eqref{eq:SA} has non-analyticities at $\zeta= \sqrt{r}$ and $\zeta= 1/\sqrt{r}$. An instance in which these two points are visible is the mid-blue solid curve of Fig.~\ref{fig:rotatedrect}. More irregular regions have many of such non-analytic points and originate peculiar looking curves like the one in Fig.~\ref{fig:star}.

\section{Conclusions}
\label{sec:conclusions}

In this paper, we studied the spreading of entanglement in a free fermionic system defined on a lattice of dimension $d\geq 1$ by generalising the quasiparticle picture of Calabrese and Cardy~\cite{quasi2005}. In particular, we showed that if the initial state has a fixed number of particles, and is invariant under no less than $\nu_j$ discrete lattice shifts in the direction $j=1,\ldots, d$, the quench produces a multiplet of $\nu_1 \cdots \nu_d$ correlated quasiparticles. This means that only settings that are effectively one-dimensional can produce pairs: in all non-degenerate cases one has to consider larger multiplets. 

Characterising the spreading of entanglement by generic multiplets of quasiparticles, we derived a general integral formula, Eq.~\eqref{eq:SAQP}, for the evolution of the entanglement entropy. Then, we studied its explicit predictions for $d=1,2$ in the case of a square lattice. In particular, we introduced an efficient Monte Carlo scheme for $d \geq 1$ to study the entanglement of \emph{arbitrary} connected regions in $d=2$.

First, we showed that exact diagonalisation results recover the generalised quasiparticle description in the limit of large subsystems and times, i.e., when the quasiparticle picture is expected to apply. Then, we studied how the entanglement dynamics depends on shape and orientation of the subsystem with respect to the underlying lattice. We observed that, for subsystems with cyclic symmetry group of order larger than three, the dependence on the orientation is negligible. Moreover, we showed that irregular regions show a multi-slope entanglement growth. Interestingly, our results also provided simple examples showing that specifying the expectation values of all conserved charges is not enough to fully determine the entanglement dynamics, even at leading order. Namely, we found examples of two different initial states with the same expectation values for all local conserved charges that generate different entanglement dynamics.  

Our general quasiparticle formulation and its Monte Carlo implementation provide a flexible and versatile method to study entanglement dynamics in non-interacting systems. The approach can directly be applied in charge conserving fermionic systems with arbitrary dispersion relation --- including the case of anisotropic couplings --- to study the entanglement of arbitrary regions in hypercubic lattices of generic dimensions $d>1$. However, it can also be directly generalised to study systems without charge conservation, e.g. BCS-like ones, and to systems on arbitrary (regular) lattices. In all these cases, whenever the initial state is not one-site shift invariant, the entanglement will generically be transported by $n$-plets of correlated quasiparticles quasiparticles with $n>2$.  

%opens up immensely the scope to study systems with qualitatively distinct, multi-slope entanglement growth, as well as to understand the analytic origin of these multiple slopes. In relation to this aspect, we remark that that distinct entanglement growth, observed here as a behaviour of cyclic group of a subsystem on a square lattice, would also be expected for the inverse case in which one considers a hypercubic system with different lattice symmetries. In this case, one must adapt the definition of the Fourier transform (\ref{eq:modes}) to the chosen lattice structure, giving rise to more finely structured quasiparticle dynamics and thus more areas $\mathcal{A}_a$. Consequently, the number of non-analycities and, in particular, those which are visible, is expected to be greater, leading to a characteristic, multi-slope entanglement growth.

%In fact, it can be directly generalised to arbitrary lattices with a suitable definition of the Fourier transform. This offers a possible direction for this work, allowing one to gain an intuition into the qualitative behaviour of entanglement growth with respect to both the subsystem and underlying lattice. We also expect that, when the lattice is promoted to three dimensions and the subsystem becomes classified by an additional cyclic symmetry group, more interesting and complicated behaviour similarly to be observed. The general treatment in this work to $d \geq 1$ dimensions would again allow such cases to be explored quite freely. 

A more immediate future direction for our work, however, is for us to use our generalised quasiparticle description to study the restoration of a discrete symmetry broken by the initial state and the possible occurrence of the quantum Mpemba effect~\cite{ares2022entanglement, rylands2023microscopic}. This can be efficiently done by using the recently introduced entanglement asymmetry~\cite{ares2022entanglement,ferro2023nonequilibrium}, which can be treated using the quasiparticle picture.  

%\section*{Note added}

{\it Note Added.} While this manuscript was being finalised, we became aware of the related work~\cite{yamashika2023time}. The latter also studies entanglement dynamics in higher dimensional free fermionic systems (${d=2}$) but focuses on special regions that can be treated by the technique of dimensional reduction~\cite{chung2000density} (see also Refs.~\cite{ares2014excited, murciano2020symmetry}).

\begin{acknowledgments}
We thank the authors of Ref.~\cite{yamashika2023time} for sharing their preliminary draft with us. This work has been supported by the Royal Society through the University Research Fellowship No.\ 201101 (B.\ B.). B.\ B.\ warmly acknowledges the hospitality of the Simons Center for Geometry and Physics during the program ``Fluctuations, Entanglements, and Chaos: Exact Results" where this work has been completed.
A.\ S.\ and A.\ J.\ acknowledge support from a research fellowship from the The Royal Commission for the Exhibition of 1851. 
\end{acknowledgments}

\appendix

\section{Numerical Solution}
\label{app:numerics}

The standard approach to obtain the von Neumann entropy of the subsystem is to replace the correlation matrix of Eq. \eqref{contributions} with the position space correlation matrix $C_A$ of the subsystem. We take the thermodynamic limit to provide the most accurate comparison with our quasiparticle solution. This also allows us to solve the elements of the full system correlation matrix in integral form so that we may construct $C_A$ directly without being limited by system size. First finding the initial correlations in Fourier space

\begin{widetext}
\begin{align*}
    \bra{\psi^{\myvec{\nu}}}c_{\myvec{p}}^\dag c_{\raisebox{-1.2pt}{$\myvec{\scriptstyle p'}$}}\ket{\psi^{\myvec{\nu}}}  &= \frac{1}{|\myvec{L}|}\sum_{\myvec{n},\myvec{m}=1}^{\myvec{L}}\bra{\psi}c^{\dag}_{\myvec{n}} c_{\myvec{m}}^{\vphantom{\dag}}\ket{\psi}e^{-i(\myvec{n}\cdot\myvec{p}-\myvec{m}\cdot\myvec{p'})}
    = \frac{1}{|\myvec{L}|}\sum_{\myvec{n},\myvec{m}=1}^{\myvec{L}} \left[ \sum_{p=1}^N   \sum_{\myvec{j}=\myvec{1}}^ {\myvec{L}/\myvec{\nu}} \delta_{\myvec{n},\myvec{m}}\delta_{\myvec{n},\myvec{\nu}\myvec{j} - \myvec{a}_p} \right] e^{-i(\myvec{n}\cdot\myvec{p}-\myvec{m}\cdot\myvec{p'})} \\
    &=\frac{1}{|\myvec{\nu}|} \left[ \frac{1}{|\myvec{L}/\myvec{\nu}|} \sum_{p=1}^N \sum_{\myvec{j}=\myvec{1}}^{\myvec{L}/\myvec{\nu}} e^{i(\myvec{p'}-\myvec{p})\cdot(\myvec{\nu} \myvec{j} - \myvec{a}_p)} \right]
    = \frac{1}{|\myvec{\nu}|}  \sum_{p=1}^N  e^{i(\myvec{p}-\myvec{p'})\cdot \myvec{a}_p} \sum_{\myvec{k}\in \frac{2\pi}{\myvec \nu}\mathbb Z_{\myvec{\nu}}} \delta_{\myvec{p},\myvec{p'}+\myvec{k}}. \numberthis
\end{align*}

\newpage

This gives the time-dependent correlations

\begin{align*}
    \bra{\psi^{\myvec{\nu}}}c_{\myvec{n}}^\dag(t)c_{\myvec{m}}(t)\ket{\psi^{\myvec{\nu}}}
    &= \frac{1}{|\myvec{L}|}\sum_{\myvec{p},\myvec{p'}\in \frac{2\pi}{\myvec L}\mathbb Z_{\myvec{L}/\myvec{\nu}}}  e^{i(\myvec{n}\cdot\myvec{p}-\myvec{m}\cdot\myvec{p'})} e^{i(\epsilon(\myvec{p'})-\epsilon(\myvec{p}))t} \bra{\psi^{\myvec{\nu}}}c_{\myvec{p}}^\dag c_{\raisebox{-1.2pt}{$\myvec{\scriptstyle p'}$}}\ket{\psi^{\myvec{\nu}}} \\
    &= \frac{1}{|\myvec{\nu L}|}\sum_{p=1}^N \sum_{\myvec{p},\myvec{p'}\in \frac{2\pi}{\myvec L}\mathbb Z_{\myvec{L}/\myvec{\nu}}}   e^{i(\myvec{n}\cdot\myvec{p}-\myvec{m}\cdot\myvec{p'})} e^{i(\epsilon(\myvec{p'})-\epsilon(\myvec{p}))t}  \left[ \sum_{p=1}^N e^{i(\myvec{p}-\myvec{p'})\cdot\myvec{a}_p} \sum_{\myvec{k}\in \frac{2\pi}{\myvec \nu}\mathbb Z_{\myvec{\nu}}} \delta_{\myvec{p},\myvec{p'}+\myvec{k}} \right] \\
    &= \frac{1}{|\myvec{\nu L}|} \sum_{p=1}^N \sum_{\myvec{p}\in \frac{2\pi}{\myvec L}\mathbb Z_{\myvec{L}/\myvec{\nu}}} \sum_{\myvec{k}\in \frac{2\pi}{\myvec \nu}\mathbb Z_{\myvec{\nu}}} e^{i\myvec{p}\cdot(\myvec{n}-\myvec{m})}
    e^{i \myvec{k}\cdot(\myvec{m}+\myvec{a}_p)} e^{it(\epsilon(\myvec{p}-\myvec{k}) -\epsilon(\myvec{p}))}. \numberthis
\end{align*}    
In the thermodynamic limit $L\to\infty$ this becomes
\be
   \lim_{{L}\to \infty} \bra{\psi^{\myvec{\nu}}}c_{\myvec{n}}^\dag(t)c_{\myvec{m}}(t)\ket{\psi^{\myvec{\nu}}} = \frac{1}{|\myvec{\nu}|} \sum_{p=1}^N \int_{0}^{2\pi} \frac{\myvec{{\rm d}p}}{\myvec{2\pi}} \sum_{\myvec{k}\in \frac{2\pi}{\myvec \nu}\mathbb Z_{\myvec{\nu}}}  e^{i\myvec{p}\cdot(\myvec{n}-\myvec{m})}
    e^{i \myvec{k}\cdot(\myvec{m}+\myvec{a}_p)} e^{it(\epsilon(\myvec{p}-\myvec{k}) -\epsilon(\myvec{p}))}. 
\ee
\end{widetext}

\section{Fourier Transform of Classical Configurations}
\label{fourier transform}

The initial state defined by Eq. \eqref{init} may be written as

\begin{equation}
\label{eq:initclassical}
    \ket*{\tilde {\psi}_{\myvec{\nu}}} = \prod_{p=1}^N b^{\dag}_{\myvec{\nu},p} \ket{0} \quad ; \quad
    b^\dag_{\myvec{\nu},p} = \prod_{\myvec{j}=\myvec{1}}^{\myvec{L} / \myvec{\nu}} c^\dag_{\myvec{\nu} \myvec{j} - \myvec{a}_p}
\end{equation}

The Fourier transform is defined for a square lattice of spatial dimension $D$ and length $\myvec{L}_d$ along each spatial dimension. Provided each $\myvec{L}_d / \myvec{\nu}_d$ is an integer, the Fourier transform of $c^\dag_{\myvec{\nu}\cdot \myvec{j} - \myvec{a}_p}$ may be written as

\begin{align*}
\label{eq:fourierdag}
    c^\dag_{\myvec{\nu}\myvec{j}-\myvec{a}_p} &=   \frac{1}{|\myvec{\nu}|^{1/2}} \sum_{\myvec{p}\in \frac{2\pi}{\myvec L}\mathbb Z_{\myvec{L}/\myvec{\nu}}} \sum_{\myvec{k}\in \frac{2\pi}{\myvec \nu}\mathbb Z_{\myvec{\nu}}} e^{i(\myvec{p}+ \myvec{k})\cdot(\myvec{\nu}\myvec{j}-\myvec{a}_p)} \tilde{c}^\dag_{\myvec{p}+\myvec{k}} \\
    &= \hspace{-5pt} \sum_{\myvec{p}\in \frac{2\pi}{\myvec L}\mathbb Z_{\myvec{L}/\myvec{\nu}}} \hspace{-5pt} e^{i\myvec{p}\cdot(\myvec{\nu}\myvec{j}-\myvec{a}_p)} \Bigg( \frac{1}{|\myvec{\nu}|^{1/2}} \sum_{\myvec{k}\in \frac{2\pi}{\myvec \nu}\mathbb Z_{\myvec{\nu}}} e^{-i\myvec{k}\cdot\myvec{a}_p} \tilde{c}^\dag_{\myvec{p}+\myvec{k}}\Bigg) \\
    &= \hspace{-5pt} \sum_{\myvec{p}\in \frac{2\pi}{\myvec L}\mathbb Z_{\myvec{L}/\myvec{\nu}}} \hspace{-5pt} e^{i\myvec{p}\cdot(\myvec{\nu}\myvec{j}-\myvec{a}_p)} B^\dag_{\myvec{\nu},\myvec{p}} \numberthis
\end{align*}

Applying Eq. \eqref{eq:fourierdag} to the operator $b^\dag_{\myvec{\nu},p}$ of Eq. \eqref{eq:initclassical} gives the Fourier transform of this operator as

\begin{align*}
    &b^{\dag}_{\myvec{\nu},p} = \prod_{\myvec{j}=1}^{\myvec{L} / \myvec{\nu}} \Bigg( \sum_{\myvec{p}\in \frac{2\pi}{\myvec L}\mathbb Z_{\myvec{L}/\myvec{\nu}}} e^{i\myvec{p}\cdot(\myvec{\nu}\myvec{j}-\myvec{a}_p)} B^\dag_{\myvec{\nu},\myvec{p}} \Bigg) \\ &= \prod_d \Bigg[ \sum_{\{\myvec{\sigma}\}} \sign(\myvec{\sigma})
    \Bigg(\prod_{n=1}^{\myvec{L}_d/\myvec{\nu}_d} e^{i \myvec{p}_{\sigma_n} (\myvec{\nu}_d n - \myvec{a}_p)}\Bigg) \Bigg] 
    \Bigg(\prod_{\myvec{p}} B^\dag_{\myvec{\nu},\myvec{p}} \Bigg) \\
    &=\Bigg( \prod_d \left[ \det( e^{i \myvec{p}_{\alpha} (\myvec{\nu}_d \beta - \myvec{a}_p)})\right]_{\alpha,\beta \in \mathbb Z_{\myvec{L}_d/\myvec{\nu}_d}}\Bigg) \Bigg(\prod_{\myvec{p}} B^\dag_{\myvec{\nu},\myvec{p}} \Bigg) \numberthis
\end{align*}

where $(B_{\myvec{\nu},\myvec{p}}^{\dag})^2 = 0$ is used to write as a sum over all possible permutations of $n$. Using that $e^{i \myvec{p}_{\alpha} (\myvec{\nu}_d \beta - \myvec{a}_p)}$ is a Van der Monde matrix, we can then write its determinant as \cite{vandermonde}

\begin{equation}
    \det( e^{i \myvec{p}_{\alpha} (\myvec{\nu}_d \beta - \myvec{a}_p)}) = \hspace{-10pt} \prod_{0\leq \alpha<\beta\leq \myvec{L}_d/\myvec{\nu}_d} \hspace{-10pt} e^{-i\myvec{a}_p}\left(e^{i \myvec{\nu}_d \alpha} - e^{i \myvec{\nu}_d \beta}\right)
\end{equation}

Since this difference of elements is nonzero for all $\alpha \neq \beta$, we see that the determinant of this matrix is also nonzero. The normalisation of $B^\dag_{\myvec{\nu},\myvec{p}}$ then implies that this determinant must be equal to one. We therefore have

\begin{align*}
\label{fourier operator}
    b^{\dag}_{\myvec{\nu},p} = \hspace{-10pt} \prod_{\myvec{p}\in \frac{2\pi\nu}{\myvec L}\mathbb Z_{\myvec{L}/\myvec{\nu}}} \hspace{-10pt} B^\dag_{\myvec{\nu},\myvec{p}} = \frac{1}{|\myvec{\nu}|^{1/2}}  \prod_{\myvec{p}\in \frac{2\pi}{\myvec L}\mathbb Z_{\myvec{L}/\myvec{\nu}}} \sum_{\myvec{k}\in \frac{2\pi}{\myvec \nu}\mathbb Z_{\myvec{\nu}}} e^{-i\myvec{k}\cdot\myvec{a}_p} \tilde{c}^\dag_{\myvec{p}+\myvec{k}} \numberthis
\end{align*}

Applying this result to the initial state of Eq. \eqref{eq:initclassical} leads to the Fourier transform of this state given by Eq. \eqref{eq:initfourier},

\begin{align*}
    \ket*{\tilde {\psi}_{\myvec{\nu}}} &= \prod_{\myvec{p}\in \frac{2\pi}{\myvec L}\mathbb Z_{\myvec{L}/\myvec{\nu}}} \Bigg( \frac{1}{|\myvec{\nu}|^{1/2}}  \prod_{p=1}^N  \sum_{\myvec{k}\in \frac{2\pi}{\myvec \nu}\mathbb Z_{\myvec{\nu}}} e^{-i {\myvec{k}} \cdot \myvec{a}_p} \tilde{c}^{\dag}_{\myvec{p}+\myvec{k}} \Bigg) \\
    & = \oprod_{\myvec{p}\in \frac{2\pi}{\myvec L}\mathbb Z_{\myvec{L}/\myvec{\nu}}} \ket*{\tilde {\psi}_{\myvec{\nu},\myvec{p}}}.
\end{align*}

{

\section{Fourier Transform of General Initial States}
\label{gaussian}

Given that the initial state \eqref{eq:initgeneral} is Gaussian, its density matrix has the exponential form

\begin{align*}
    \hat{\rho} &= \prod_{\myvec{j}=1}^{\myvec{\nu}} \sum_{\myvec{n},\myvec{m}=1}^{\myvec{\nu}} \exp\bigg[ c^{\dag}_{\myvec{\nu j} + \myvec{n}} A_{\myvec{n},\myvec{m}}^{\vphantom{\dag}} c_{\myvec{\nu j} + \myvec{m}}^{\vphantom{\dag}} \bigg] \\ 
    &= \exp\bigg[ \sum_{\myvec{j}=1}^{\myvec{L}/\myvec{\nu}} \sum_{\myvec{n},\myvec{m}=1}^{\myvec{\nu}} c^{\dag}_{\myvec{\nu j} + \myvec{n}} A_{\myvec{n},\myvec{m}}^{\vphantom{\dag}} c_{\myvec{\nu j} + \myvec{m}}^{\vphantom{\dag}} \bigg] \numberthis
\intertext{The Fourier transform of these operators gives}
    & = \exp\bigg[ \sum_{\myvec{p}\in \frac{2\pi}{\myvec L}\mathbb Z_{\myvec{L}/\myvec{\nu}}} \sum_{\myvec{k},\myvec{k'}\in \frac{2\pi}{\myvec \nu}\mathbb Z_{\myvec{\nu}}}  \tilde{c}^{\dag}_{\myvec{p} + \myvec{k}} \tilde{A}_{\myvec{k}, \myvec{k'}}^{(\myvec{p})\vphantom{\dag}} \tilde{c}_{\myvec{p} + \myvec{k'}}^{\vphantom{\dag}} \bigg] \\
    & = \prod_{\myvec{p}\in \frac{2\pi}{\myvec L}\mathbb Z_{\myvec{L}/\myvec{\nu}}} \sum_{\myvec{k},\myvec{k'}\in \frac{2\pi}{\myvec \nu}\mathbb Z_{\myvec{\nu}}}  \exp\bigg[ \tilde{c}^{\dag}_{\myvec{p} + \myvec{k}} \tilde{A}_{\myvec{k},\myvec{k'}}^{(\myvec{p})\vphantom{\dag}} \tilde{c}_{\myvec{p} + \myvec{k'}}^{\vphantom{\dag}} \bigg] \numberthis
\intertext{such that the density matrix takes the form}
     \hat{\rho} & = \ket{\psi_{\myvec{\nu}}} \bra{\psi_{\myvec{\nu}}} = \oprod_{\myvec{p}\in \frac{2\pi}{L}\mathbb Z_{\myvec L/\myvec \nu}} \ket*{\tilde{\psi}_{\myvec{\nu},\myvec{p}}}\bra*{\tilde{\psi}_{\myvec{\nu},\myvec{p}}}
\end{align*}

where $\ket*{\tilde{\psi} _{\myvec{\nu} ,\myvec{p}}}$ is a Gaussian state with generalised $\myvec{p}$ dependence conferred by the matrix $\tilde{A}^{(\myvec{p})}$. This gives the Fourier transform of the initial state given by Eq.~\eqref{eq:initgeneral}.
}

\section{Monte Carlo Scheme}
\label{app:MC}

Here we outline the Monte Carlo scheme to solve Eq.~\eqref{eq:SAQP}. 

The quasiparticle solution to the entanglement dynamics can be written as the integral
\begin{equation}
    \!\!\!\!\frac{S_A (t/L)}{|A|} \rightarrow  \frac{1}{|A|}\int_{\mathbb R^d} \!\!\!\!{\rm d}\myvec {x} \int_{\myvec{0}}^{\frac{\myvec{2\pi}}{\myvec{\nu}}} \!\!\!\frac{{\rm d} \myvec {p}}{\myvec{2\pi}} S(\rho_{A}(\myvec{p},\myvec{x},t/L)),
    \label{eq: Monte Carlo integral}
\end{equation}
in the limit $t, |A| \rightarrow \infty$ with $t/|A|=\text{fixed}$, $L$ is a linear dimension, and where the integral is over the particle entanglement of all multiplets at a given time $t$. We can numerically evaluate this integral using a Monte Carlo scheme that mimics the classical dynamics of the modes in the multiplets and sums the corresponding entanglement contributions. Note that sampling over all position space would not be efficient as most points would not contribute to the entanglement, so instead we can sample all points inside the subsystem at time $t$ and evolve these points backward to effectively sample within a region that may contribute to the entanglement, so long as we correctly account for any possible double counting. 

The Monte Carlo integration proceeds as follows:
\begin{enumerate}
\item Generate a random position $\tilde{\myvec{x}}$, sampled from a uniform distribution over the region $A$. For an irregular region $A$, this can most easily be done by uniformly sampling from a rectangular region that bounds $A$ and only accept the sample if the point is in $A$. If the sample is not in $A$, then we continue sampling until we get one that is. This will correspond to the position of a selected mode in the multiplet at time $t$.

\begin{figure}[t]
\centering
        \begin{adjustbox}{width=0.4\textwidth}
            \includegraphics{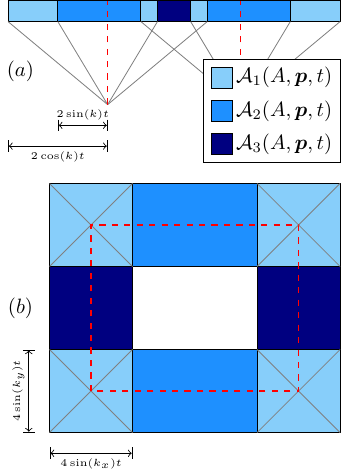}%
        \end{adjustbox}
        \caption{Diagram of the quasiparticle dynamics for a) a $\nu=4$ initial state, and b) $\nu_x=\nu_y=2$ initial state and rectangular subsystem with $l_x>l_y$. The red dashed line marks the boundary of the subsystem and the three areas $\{{\cal A}_j(A, \bm{p},t)\}_{j=1,2,3}$ are depicted for a) $0<p<\pi/4$ and $l/2(\sin(p)+\cos(p))<t<l/4\sin(p)$, and b) $p_x=p_y$ and $t<l_x/4\sin(p_x)$. These areas are obtained by tracing the motion of quasiparticles as outlined in Section \ref{sec:qppicture}, and their explicit solutions are given in Appendix \ref{app:dynamics}.}
        \label{appfig:areas}
\end{figure}

\item Generate a random momentum $\myvec{p}\in [\myvec{0}, 2\pi/\myvec{\nu}]$, and randomly chose one of the $\nu$ modes in the multiplet, labelled by $\myvec{n}$. For example, in $d=2$, we select the mode labelled by $n_x \in \{0,\ldots, \nu_x-1\}$ and $n_y = \{0,\ldots, \nu_y-1\}$.
\item Evolve the mode back to time $t=0$. That is, find $\myvec{x} = \tilde{\myvec{x}} + 2J\sin(\myvec{p} + \frac{\myvec{n}}{\myvec{\nu}} 2\pi ) t$. 
\item We now have an initial $\myvec{x}$, and $\myvec{p}$. We then evolve all of the modes in the multiplet forward in time to find their position at time $t$. For the mode labelled by $\myvec{n}_i$, that is $\myvec{x}_i(t) = \myvec{x} - 2J\sin(\myvec{p} + \frac{\myvec{n}_i}{\myvec{\nu}} 2\pi ) t$.
\item Given the positions of all the modes at time $t$, we then note which are inside the region $A$ and compute the corresponding particle entanglement contribution. This generally depends on which particles are in the region, as well as on $\myvec{p}$. Finally, we divide this entanglement contribution by the number of modes that are in the region $A$, and then add this to the sum. Dividing by the number of modes inside $A$ corrects for the over-counting of different $\tilde{\myvec{x}}$ and $\myvec{n}$ that correspond to the same multiplet.
\end{enumerate}

In the end, we are left with a sum over all particle entanglement contributions, which we divide by the number of samples taken to get the approximation of the integral Eq.~\eqref{eq: Monte Carlo integral}. Note that, by also computing the average of the squares of the particle entanglement contributions, we can also keep track of the variance $\sigma^2$ of the Monte Carlo sampling, and so estimate the standard error of the mean $\sigma/\sqrt{\text{number of samples}}$.

\section{Explicit form of ${\cal A}_j$ for rectangular regions in $d=1,2$}
\label{app:dynamics}

Here, we present the analytic solutions to the functions ${\cal A}_j(A,\myvec{p},t)$ defined by Eq.~\ref{eq:xintegratedqp}. Following sections~\ref{sec:resultsbasic} and~\ref{sec:rotate}, we focus on $d=1$ and $d=2$ initial states, where $A$ is a simple hypercubic subsystem.
\\

We begin with the $d=1$ states of \ref{sec:resultsbasic}, namely with $\nu=4$ states in a subsystem of length $l$, whose three areas $\{{\cal A}_j(A,p,t)\}_{j=1,2,3}$ are specified by Fig. \ref{appfig:areas}(a). First, we write the solutions to ${\cal A}_j(A,p,t)$ for arbitrary velocity ordering $v_a>v_b>v_c>v_d$. For convenience, we define the natural times $\tau_{ab}=l/(v_a-v_b)$ and lengths $\Delta_{ab} = (v_a-v_b)t$, where we have applied this symmetry of mode velocities $v_c = -v_b$ and $v_d = -v_a$, to write \\

\begin{widetext}
\begin{align*}
    {\cal A}_1(A,p,t) =& 4\Delta_{ab}H(\tau_{ad}-t) + 2(l-\Delta_{bc})H(\tau_{ac}-t)H(t-\tau_{ad}) + 2(\Delta_{ad}-\Delta_{ac}-l)H(\min(\tau_{ab},\tau_{bc})-t) \cdot \\ & H(t-\tau_{ac}) + 2(l+\Delta_{bc})H(\tau_{bc}-t)H(t-\tau_{ab}) + 4\Delta_{ab}H(\tau_{ab}-t)H(t-\tau_{bc})+4lH(t-\max(\tau_{ab},\tau_{bc})) \\
    {\cal A}_2(A,p,t) =& 2\Delta_{bc}H(\tau_{ac}-t) + 2(l-\Delta_{ab})H(\min(\tau_{ab},\tau_{bc})-t)H(t-\tau_{ac}) + 2(l-\Delta_{ab})H(t-\tau_{bc})H(\tau_{bc}-t) \\
    {\cal A}_3(A,p,t) =& (\Delta_{ad}-l)H(t-\tau_{ad})H(\tau_{ac}-t) + (l-\Delta_{bc})H(t-\tau_{ac})(\tau_{bc}-t) \,\,\, , \numberthis
\label{eq:areasol1d}
\end{align*}
\end{widetext}

where $H(x)$ is the Heaviside step function. These solutions can be rearranged into a full timewise solution of the quasiparticle dynamics as shown in Fig.~\ref{apptbl:1d timewise}.

\begin{table}[H]
\centering
    \begin{tabular}{ R{3.6cm} | L{4.5cm} }
        \hline
        \textbf{Time Interval} & \textbf{Total Entropy $\nu = 4$} \\
        \hline
        $\myvec{t<\tau_{ad}}$ & $4\Delta_{ab}\cdot s_1 + 2\Delta_{bc}\cdot s_2$ \\
        \hline
        $\myvec{\tau_{ad} < t < \tau_{ac}}$ & $ 2(l-\Delta_{bc})\cdot s_1 + 2\Delta_{bc}\cdot s_2 + (\Delta_{ad}-l)\cdot s_3$ \\
        \hline
        $\myvec{\tau_{ac} < t < \text{min}(\tau_{ab}, \tau_{bc})}$ & $ 2(\Delta_{ab}+\Delta_{ac}-l)\cdot s_1 + 2(l-\Delta_{ab})\cdot s_2 +(l-\Delta_{bc})\cdot s_3$ \\
        \hline
        $(i)\thickspace\myvec{\tau_{ab} < t < \tau_{bc}}$ & $2  (l+\Delta_{bc})\cdot s_1 + (l-\Delta_{bc}) \cdot s_3$ \\
        \cdashline{1-2}
        $(ii)\thickspace\myvec{\tau_{bc} < t < \tau_{ab}}$ & $4 \Delta_{ab} \cdot s_1 + 2(l-\Delta_{ab})\cdot s_2$ \\
        \hline
        $\myvec{t > \text{max}(\tau_{ab}, \tau_{bc})}$ & $4l \cdot s_1 $ \\
        \hline
    \end{tabular}
    \label{apptbl:1d timewise}
\end{table}

where the contributions $\{s_j\}_{j=1,2,3}$ for the $d=1$ classical configurations of Section \ref{sec:resultsbasic} are defined in Table \ref{tbl:1d entropy contributions}. Next, for these classical configurations, we introduce the \textit{integrated areas},

\begin{equation}
  {\cal A}_j(A, t)= \int_{0}^{{\myvec{\pi}}}  \!\!\!\frac{{\rm d} \myvec {p}}{\myvec{2\pi}} {\cal A}_j(A,\bm{p},t) \,\,\, ,
  \label{eq:integratedareas}
\end{equation}

and then divide the momentum integral of Eq.~\eqref{eq:integratedareas} into intervals for which the ordering of mode velocities is fixed,

\begin{equation}
      {\cal A}_j(A,t)= \left( \int_{0}^{\pi/4}  \frac{{\rm d} p}{2\pi} + \int_{\pi/4}^{\pi/2}  \frac{{\rm d} p}{2\pi} \right) {\cal A}_j(A,p,t) \,\,\, .
\label{eq:intsplit1d}
\end{equation}

We then combine Eqs.~\eqref{eq:areasol1d} and~\eqref{eq:intsplit1d} to obtain the solution to the integrated areas $\{{\cal A}_j(A,t)\}_{j=1,2,3}$. It should be noted that, while these integrated areas yield the solution to all states of fixed $s_j(\myvec{p})=s_j$, Eq. \eqref{eq:integratedareas} can easily be modified to a state dependent solution of general $s_j(\myvec{p})$. \\

We now turn to the $d=2$ states of \ref{sec:resultsbasic}, namely to $\nu_x = \nu_y = 2$ states in a rectangular subsystem of lengths $l_x, l_y$ aligned with the lattice, whose three areas $\{{\cal A}_j(A,\myvec{p},t)\}_{j=1,2,3}$ are specified by Fig. \ref{appfig:areas} (b). Our method closely follows the $d=1$ case, where first we write the solutions to ${\cal A}_j(A,\bm{p},t)$ for arbitrary velocity ordering $v_a>v_b>v_c>v_d$, ordered by projection onto the positive $x$-axis. For convenience, we define the natural times $\tau_i \equiv l_i/4\sin(k_i)$ and lengths $X,Y \equiv 4\sin(k_i)t$ for $i = x,y$, where we have applied this symmetry of mode velocities $v_c = -v_a$ and $v_d = -v_b$, to write

\begin{widetext}
\begin{align*}
    {\cal A}_1(A,\myvec{p},t) & = 4XY H(\min(\tau_x,\tau_y)-t) + 4l_x Y H(t-\tau_x) H(\tau_y-t) + 4Xl_y H(t-\tau_y)H(\tau_x-t) + 4l_x l_y (H\max(\tau_x,\tau_y,t)) \\
    {\cal A}_2(A,\myvec{p},t) & = 2(l_x-X)Y H(\min(\tau_x,\tau_y)-t) + 2(l_x-X)l_y H(t-\tau_y)H(\tau_x-t) \\
    {\cal A}_3(A,\myvec{p},t) & = 2(l_y-Y)X H(\min(\tau_x,\tau_y)-t) + 2(l_y-Y)l_x H(t-\tau_x)H(\tau_y-t) \,\,\,. \numberthis
\label{eq:areasol2d}
\end{align*}
\end{widetext}

\begin{figure}[t]
\centering
    \includegraphics{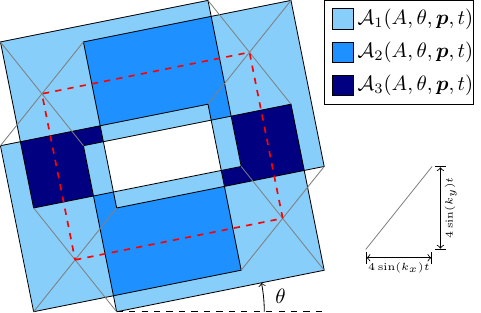}%
    \caption{
    Diagram of the quasiparticle dynamics for a $\nu_x=\nu_y=2$ initial state and rotated rectangular subsystem with $l_x>l_y$. The red dashed line marks the boundary of the subsystem and the three areas ${\cal A}_j(A, \theta, \bm{p},t)$, $j=1,2,3$, are depicted for fixed $(\bm{p}, \theta, t)$. In this case, the structure of each of these shaded regions changes with angle of rotation $\theta$ such that the full solutions to ${\cal A}_j(A, \theta, \bm{p},t)$ $j=1,\ldots,3$ are piecewise functions in $\myvec{p}$. In particular, this diagram displays the structure for all $\myvec{p}$ such that $\sin(p_y)/\sin(p_x) \in [\theta,\pi/2-\theta]$.
    \label{appfig:areasrotate}}
\end{figure}  

These solutions can be again rearranged into a full timewise solution of the quasiparticle dynamics as shown in Fig.~\ref{apptbl:2d timewise}.

\begin{table}[H]
\centering
    \begin{tabular}{ R{3.6cm} | L{4.5cm} }
        \hline
        \textbf{Time Interval} & \textbf{Total Entropy $\nu_x = \nu_y = 2$} \\
        \hline
        $\myvec{t<\text{min}(\tau_x, \tau_y)}$ & $4XY\cdot s_1 + 2(l_x-X)Y\cdot s_2 + 2(l_y-Y)X\cdot s_3 $ \\
        \hline
        $(i)\thickspace\myvec{\tau_x<t<\tau_y}$ & $4l_x Y \cdot s_1 + 2(l_y-Y)l_x \cdot s_3$ \\
        \cdashline{1-2}
        $(ii)\thickspace\myvec{\tau_y<t<\tau_x}$ & $ 4X l_y \cdot s_1 + 2(l_x-X)l_y \cdot s_2$ \\
        \hline
        $\myvec{t>\text{max}(\tau_x, \tau_y)}$ & $4 l_x l_y \cdot s_1$ \\
        \hline
    \end{tabular}
    \label{apptbl:2d timewise}
\end{table}

where the contributions $\{s_j\}_{j=1,2,3}$ for the $d=2$ classical configurations of Section \ref{sec:resultsbasic} are defined in Table \ref{tbl:2d entropy contributions}.

Then, as before, we divide the momentum integral of Eq. \eqref{eq:integratedareas} into intervals for which the ordering of mode velocities is fixed,

\begin{align*}
      {\cal A}_j(A,t) =& \left( \int_{0}^{\pi/2} \frac{{\rm d} p_x}{2\pi} + \int_{\pi/2}^{\pi}  \frac{{\rm d} p_x}{2\pi} \right) \cdot \\
      & \left( \int_{0}^{\pi/2}  \frac{{\rm d} p_y}{2\pi} + \int_{\pi/2}^{\pi}  \frac{{\rm d} p_y}{2\pi} \right) {\cal A}_j(A,\myvec{p},t) \,\,\,, \numberthis
\label{eq:intsplit2d}
\end{align*}

and combine Eqs. \eqref{eq:areasol2d} and \eqref{eq:intsplit2d} to obtain the solution to the areas $\{{\cal A}_j\}_{j=1,2,3}$.

Finally, we generalise \eqref{eq:areasol2d} to the case of Section \ref{sec:rotate}, namely to $\nu_x = \nu_y = 2$ states in a rectangular subsystem of lengths $l_x, l_y$ and angle of rotation $\theta$, whose three areas $\{{\cal A}_j\}_{j=1,2,3}$ are specified by Fig. \ref{appfig:areasrotate}.

To formulate the solution to $\{{\cal A}_j\}_{j=1,2,3}$, we first define the functions,

\begin{align*}
    \psi(k_1,k_2) &= \arctan(\sin(k_1)/\sin(k_2)) \\
    v(k_1, k_2) &= \sqrt{\sin^2(k_1) + \sin^2(k_2)} \\
    G(x) &= x \cdot H(x), \numberthis
    \label{eq:def1}
\end{align*}

from which we construct a set of functions for the lengths between various corners and points of intersection of the light cones for each mode,

\begin{align*}
    a(\myvec{k},\theta,t) &= G(1-2t\sin(k_x)\cos(\theta)) \\
    b(\myvec{k},\theta,t) &= G(1-2t\sin(k_x)\sin(\theta)) \\
    c(\myvec{k},\theta,t) &= G(1-2t\sin(k_y)\cos(\theta)) \\
    d(\myvec{k},\theta,t) &= G(1-2t\sin(k_y)\sin(\theta)) \\
    e(\myvec{k},\theta,t) &= G(1-2vt\cos(\psi(k_y,k_x)-\theta)) \\
    f(\myvec{k},\theta,t) &= G(1-2vt|\sin(\psi(k_y,k_x)-\theta)|) \\
    g(\myvec{k},\theta,t) &= G(1-2vt\cos(\psi(k_x,k_y)-\theta)) \\
    h(\myvec{k},\theta,t) &= G(1-2vt|\sin(\psi(k_x,k_y)-\theta)|) \,\,\,. \numberthis
    \label{eq:def2}
\end{align*}

These light cones are shown in Fig. \ref{appfig:areasrotate} for fixed $(\myvec{k},t)$. Finally, using these definitions \eqref{eq:def1} and \eqref{eq:def2}, we have

\begin{widetext}
\begin{align*}
    {\cal A}_1(A, \theta, \bm{p},t) &= 2e(1-f)H(\sin(\psi(k_y,k_x)-\theta)) + g(1-h)H(-\sin(\psi(k_y,k_x)-\theta)) + (a-e)(b-g) \\
    {\cal A}_2(A, \theta, \bm{p},t) &= 2e(1-f)H(-\sin(\psi(k_y,k_x)-\theta)) + g(1-h)H(\sin(\psi(k_y,k_x)-\theta)) + (c-g)(d-e) \\
    {\cal A}_3(A, \theta, \bm{p},t) &= 4(1 - ab - cd) +2(ef + gh) 
\end{align*}
\end{widetext}

Strictly, this solution is valid for $0 < \theta < \pi/4$. However, we note that any angle of rotation can be achieved from this solution by redefining the initial state.

\section{Entanglement Contributions For Superposition States}
\label{app:contributions}

In this appendix  we report the entanglement contributions for some superposition states in $d=1$ and $d=2$ which we computed via Eq.~\eqref{eq:multipletcorrelationmatrix}. 

In $d=1$ we consider a superposition state of the form
\begin{equation}\label{eq:phi-sup-1D}
\ket{\phi_{4}} \propto (\stateone{0,1}+\alpha\stateone{0,2})^{\otimes L/4},  
\end{equation}
where we omitted an overall constant ensuring normalisation. The contribution to the entanglement when only one mode of the multiplet is in or out of the system depends on the mode. Specifically we have
\begin{equation}
\begin{aligned}
 s_{p}(k)=&-\frac{1-f_p(k)}{2}\ln{(\frac{1-f_p(k)}{2})}\\
 &\quad-\frac{1+f_p(k)}{2}\ln{(\frac{1+f_p(k)}{2})},
\end{aligned}
\end{equation}
where
\be
f_p(k)=\frac{\alpha}{1+\alpha^2}\cos(k - \frac{2\pi}{\nu} p),
\ee
and $p=0,\cdots,3$ identifies the mode. Instead, when only two modes are in the system we have 
\be
\begin{aligned}
s_{p,q}(k)=&-\lambda_{p,q}(k)\log\lambda_{p,q}(k)\\
&- (1-\lambda_{p,q}(k))\log(1-\lambda_{p,q}(k)),
\end{aligned}
\ee
with
\begin{equation}
\begin{aligned}
 &\lambda_{p,q}=\frac{1}{2}\Bigg(1+\frac{f_{p}(k)}{2}+\frac{f_{q}(k)}{2}
 \pm \Bigg[\tfrac{|\sin((p-q)\tfrac{2\pi}{\nu})|}{2*(1+\alpha^2)} \\
 &+ \tfrac{\alpha^2|\cos((p-q)\tfrac{2\pi}{\nu})|}{(1+\alpha^2)}+\left(\frac{f_{p}(k)}{2}+\frac{f_{q}(k)}{2}\right)^2\Bigg]^{1/2}\Bigg).
\end{aligned}
\end{equation}
%where $g_{\alpha,p}(k)=\frac{\alpha \cos(k +\frac{2 \pi p}{\nu})}{2(1+\alpha^2)}$. 

In $d=2$ we consider the following superposition state
\begin{equation}
\ket{\phi_{2,2}} \propto \left(\statetwo{0,2}+\alpha \statetwo{0,3}\right)^{\otimes L^2/4}.
\label{eq:phi-sup-2D}
\end{equation}
The contribution of a single mode in (or out of) the system turns out to be mode-independent and is given by 
\begin{equation}
\begin{aligned}
    s_{1}(\myvec k) =&-\left(\frac{1-g(k_x)}{2}\right)\ln{\left(\frac{1-g(k_x)}{2}\right)}\\
    &-\left(\frac{1+g(k_x)}{2}\right)\ln{\left(\frac{1+ g(k_x)}{2}\right)}.
\end{aligned}
\end{equation}
Concerning the contributions of two modes, the only ones produced by the quasiparticle dynamics are
\begin{align*}
    s_{2}(\myvec k) =&2\ln{2}-\left(1-\tfrac{1}{\sqrt{1+\alpha^2}}\right)\ln{\left(1-\tfrac{1}{\sqrt{1+\alpha^2}}\right)}\\
    &-\left(1+\tfrac{1}{\sqrt{1+\alpha^2}}\right)\ln{\left(1+\tfrac{1}{\sqrt{1+\alpha^2}}\right)},
\end{align*}
\begin{align*}
    s_{3}(\myvec k) =&\ln{2}-\left(\frac{1}{2}-g(k_x)\right)\ln{\left(\frac{1}{2}- g(k_x)\right)}\\
    &-\left(\frac{1}{2}+g(k_x)\right)\ln{\left(\frac{1}{2}+ g(k_x)\right)}, \numberthis
\end{align*}
where $g(\myvec k)=\alpha \cos(\myvec k)/(1+\alpha^2)$. The saturation value for entanglement density is given by
\begin{equation}
    S_{A,\alpha}(\infty)=\tfrac{\sqrt{\alpha^4+\alpha^2+1}}{\alpha^2+1}-1-\ln{(\tfrac{1}{4}+\tfrac{\sqrt{\alpha^4+\alpha^2+1}}{4(\alpha^2+1)})},
\end{equation}
for both the $d=1$ and $d=2$ cases. 
In the limit $\alpha\rightarrow0,\infty$ these contributions recover those presented in the second two columns of Tabs.~\ref{tbl:1d entropy contributions} and \ref{tbl:2d entropy contributions}. 

We remark that the initial state should be Gaussian in order for our techniques to apply. This property is not immediately obvious for superposition states; however, in our translational invariant setting we can verify it by restricting to the unit cell. Specifically, we require a general superposition state such as 
\begin{equation}
    \ket*{\psi^{\rm c}_{\myvec{\nu},\myvec{j}}} = \sum_{\myvec{\ell}_i \in \mathbb Z_{\myvec{\nu}}} \alpha_{\myvec{\ell}} c^\dag_{\myvec{\ell}_1+ \myvec{\nu} \myvec{j}} \hdots c^\dag_{\myvec{\ell}_N+\myvec{\nu} \myvec{j}} \ket{0},
\end{equation}
to be annihilated by a new family of canonical fermions $f_{\myvec{i}}$ that is linearly related to $c_{\myvec{i}}$ and $c^\dag_{\myvec{i}}$. Namely, one can write 
\begin{equation}
\ket*{\psi^{\rm c}_{\myvec{\nu},\myvec{j}}} = \prod_{\myvec{i}}f^\dagger_{\myvec{i}} \ket{0}.
\label{eq:gaussianity}
\end{equation}
This is always the case for the superpositions considered here.

\bibliography{refs}

\begin{thebibliography}{69}%
\makeatletter
\providecommand \@ifxundefined [1]{%
 \@ifx{#1\undefined}
}%
\providecommand \@ifnum [1]{%
 \ifnum #1\expandafter \@firstoftwo
 \else \expandafter \@secondoftwo
 \fi
}%
\providecommand \@ifx [1]{%
 \ifx #1\expandafter \@firstoftwo
 \else \expandafter \@secondoftwo
 \fi
}%
\providecommand \natexlab [1]{#1}%
\providecommand \enquote  [1]{``#1''}%
\providecommand \bibnamefont  [1]{#1}%
\providecommand \bibfnamefont [1]{#1}%
\providecommand \citenamefont [1]{#1}%
\providecommand \href@noop [0]{\@secondoftwo}%
\providecommand \href [0]{\begingroup \@sanitize@url \@href}%
\providecommand \@href[1]{\@@startlink{#1}\@@href}%
\providecommand \@@href[1]{\endgroup#1\@@endlink}%
\providecommand \@sanitize@url [0]{\catcode `\\12\catcode `\$12\catcode
  `\&12\catcode `\#12\catcode `\^12\catcode `\_12\catcode `\%12\relax}%
\providecommand \@@startlink[1]{}%
\providecommand \@@endlink[0]{}%
\providecommand \url  [0]{\begingroup\@sanitize@url \@url }%
\providecommand \@url [1]{\endgroup\@href {#1}{\urlprefix }}%
\providecommand \urlprefix  [0]{URL }%
\providecommand \Eprint [0]{\href }%
\providecommand \doibase [0]{https://doi.org/}%
\providecommand \selectlanguage [0]{\@gobble}%
\providecommand \bibinfo  [0]{\@secondoftwo}%
\providecommand \bibfield  [0]{\@secondoftwo}%
\providecommand \translation [1]{[#1]}%
\providecommand \BibitemOpen [0]{}%
\providecommand \bibitemStop [0]{}%
\providecommand \bibitemNoStop [0]{.\EOS\space}%
\providecommand \EOS [0]{\spacefactor3000\relax}%
\providecommand \BibitemShut  [1]{\csname bibitem#1\endcsname}%
\let\auto@bib@innerbib\@empty
%</preamble>
\bibitem [{\citenamefont {von Neumann}(2010)}]{von2010proof}%
  \BibitemOpen
  \bibfield  {author} {\bibinfo {author} {\bibfnamefont {J.}~\bibnamefont {von
  Neumann}},\ }\bibfield  {title} {\bibinfo {title} {Proof of the ergodic
  theorem and the h-theorem in quantum mechanics: Translation of: Beweis des
  ergodensatzes und des h-theorems in der neuen mechanik},\ }\href
  {https://doi.org/10.1140/epjh/e2010-00008-5} {\bibfield  {journal} {\bibinfo
  {journal} {The European Physical Journal H}\ }\textbf {\bibinfo {volume}
  {35}},\ \bibinfo {pages} {201} (\bibinfo {year} {2010})}\BibitemShut
  {NoStop}%
\bibitem [{\citenamefont {Calabrese}\ \emph {et~al.}(2016)\citenamefont
  {Calabrese}, \citenamefont {Essler},\ and\ \citenamefont
  {Mussardo}}]{calabrese2016introduction}%
  \BibitemOpen
  \bibfield  {author} {\bibinfo {author} {\bibfnamefont {P.}~\bibnamefont
  {Calabrese}}, \bibinfo {author} {\bibfnamefont {F.~H.~L.}\ \bibnamefont
  {Essler}},\ and\ \bibinfo {author} {\bibfnamefont {G.}~\bibnamefont
  {Mussardo}},\ }\bibfield  {title} {\bibinfo {title} {Introduction to `quantum
  integrability in out of equilibrium systems'},\ }\href
  {https://doi.org/10.1088/1742-5468/2016/06/064001} {\bibfield  {journal}
  {\bibinfo  {journal} {J. Stat. Mech. Theory Exp.}\ }\textbf {\bibinfo
  {volume} {2016}},\ \bibinfo {pages} {064001} (\bibinfo {year}
  {2016})}\BibitemShut {NoStop}%
\bibitem [{\citenamefont {Vidmar}\ and\ \citenamefont
  {Rigol}(2016)}]{vidmar2016generalized}%
  \BibitemOpen
  \bibfield  {author} {\bibinfo {author} {\bibfnamefont {L.}~\bibnamefont
  {Vidmar}}\ and\ \bibinfo {author} {\bibfnamefont {M.}~\bibnamefont {Rigol}},\
  }\bibfield  {title} {\bibinfo {title} {Generalized gibbs ensemble in
  integrable lattice models},\ }\href
  {https://doi.org/10.1088/1742-5468/2016/06/064007} {\bibfield  {journal}
  {\bibinfo  {journal} {J. Stat. Mech. Theory Exp.}\ }\textbf {\bibinfo
  {volume} {2016}},\ \bibinfo {pages} {064007} (\bibinfo {year}
  {2016})}\BibitemShut {NoStop}%
\bibitem [{\citenamefont {Essler}\ and\ \citenamefont
  {Fagotti}(2016)}]{essler2016quench}%
  \BibitemOpen
  \bibfield  {author} {\bibinfo {author} {\bibfnamefont {F.~H.~L.}\
  \bibnamefont {Essler}}\ and\ \bibinfo {author} {\bibfnamefont
  {M.}~\bibnamefont {Fagotti}},\ }\bibfield  {title} {\bibinfo {title} {Quench
  dynamics and relaxation in isolated integrable quantum spin chains},\ }\href
  {https://doi.org/10.1088/1742-5468/2016/06/064002} {\bibfield  {journal}
  {\bibinfo  {journal} {J. Stat. Mech. Theory Exp.}\ }\textbf {\bibinfo
  {volume} {2016}},\ \bibinfo {pages} {064002} (\bibinfo {year}
  {2016})}\BibitemShut {NoStop}%
\bibitem [{\citenamefont {Doyon}(2020)}]{doyon2020lecture}%
  \BibitemOpen
  \bibfield  {author} {\bibinfo {author} {\bibfnamefont {B.}~\bibnamefont
  {Doyon}},\ }\bibfield  {title} {\bibinfo {title} {{Lecture Notes On
  Generalised Hydrodynamics}},\ }\href
  {https://doi.org/10.21468/SciPostPhysLectNotes.18} {\bibfield  {journal}
  {\bibinfo  {journal} {SciPost Phys. Lect. Notes}\ ,\ \bibinfo {pages} {18}}
  (\bibinfo {year} {2020})}\BibitemShut {NoStop}%
\bibitem [{\citenamefont {Bastianello}\ \emph {et~al.}(2022)\citenamefont
  {Bastianello}, \citenamefont {Bertini}, \citenamefont {Doyon},\ and\
  \citenamefont {Vasseur}}]{bastianello2022introduction}%
  \BibitemOpen
  \bibfield  {author} {\bibinfo {author} {\bibfnamefont {A.}~\bibnamefont
  {Bastianello}}, \bibinfo {author} {\bibfnamefont {B.}~\bibnamefont
  {Bertini}}, \bibinfo {author} {\bibfnamefont {B.}~\bibnamefont {Doyon}},\
  and\ \bibinfo {author} {\bibfnamefont {R.}~\bibnamefont {Vasseur}},\
  }\bibfield  {title} {\bibinfo {title} {Introduction to the special issue on
  emergent hydrodynamics in integrable many-body systems},\ }\href
  {https://doi.org/10.1088/1742-5468/ac3e6a} {\bibfield  {journal} {\bibinfo
  {journal} {J. Stat. Mech. Theory Exp.}\ }\textbf {\bibinfo {volume} {2022}},\
  \bibinfo {pages} {014001} (\bibinfo {year} {2022})}\BibitemShut {NoStop}%
\bibitem [{\citenamefont {Alba}\ \emph {et~al.}(2021)\citenamefont {Alba},
  \citenamefont {Bertini}, \citenamefont {Fagotti}, \citenamefont {Piroli},\
  and\ \citenamefont {Ruggiero}}]{alba2021generalized}%
  \BibitemOpen
  \bibfield  {author} {\bibinfo {author} {\bibfnamefont {V.}~\bibnamefont
  {Alba}}, \bibinfo {author} {\bibfnamefont {B.}~\bibnamefont {Bertini}},
  \bibinfo {author} {\bibfnamefont {M.}~\bibnamefont {Fagotti}}, \bibinfo
  {author} {\bibfnamefont {L.}~\bibnamefont {Piroli}},\ and\ \bibinfo {author}
  {\bibfnamefont {P.}~\bibnamefont {Ruggiero}},\ }\bibfield  {title} {\bibinfo
  {title} {Generalized-hydrodynamic approach to inhomogeneous quenches:
  correlations, entanglement and quantum effects},\ }\href
  {https://doi.org/10.1088/1742-5468/ac257d} {\bibfield  {journal} {\bibinfo
  {journal} {J. Stat. Mech. Theory Exp.}\ }\textbf {\bibinfo {volume} {2021}},\
  \bibinfo {pages} {114004} (\bibinfo {year} {2021})}\BibitemShut {NoStop}%
\bibitem [{\citenamefont {Polkovnikov}\ \emph {et~al.}(2011)\citenamefont
  {Polkovnikov}, \citenamefont {Sengupta}, \citenamefont {Silva},\ and\
  \citenamefont {Vengalattore}}]{PolkovnikovReview}%
  \BibitemOpen
  \bibfield  {author} {\bibinfo {author} {\bibfnamefont {A.}~\bibnamefont
  {Polkovnikov}}, \bibinfo {author} {\bibfnamefont {K.}~\bibnamefont
  {Sengupta}}, \bibinfo {author} {\bibfnamefont {A.}~\bibnamefont {Silva}},\
  and\ \bibinfo {author} {\bibfnamefont {M.}~\bibnamefont {Vengalattore}},\
  }\bibfield  {title} {\bibinfo {title} {Colloquium: Nonequilibrium dynamics of
  closed interacting quantum systems},\ }\href
  {https://doi.org/10.1103/RevModPhys.83.863} {\bibfield  {journal} {\bibinfo
  {journal} {Rev. Mod. Phys.}\ }\textbf {\bibinfo {volume} {83}},\ \bibinfo
  {pages} {863} (\bibinfo {year} {2011})}\BibitemShut {NoStop}%
\bibitem [{\citenamefont {Gogolin}\ and\ \citenamefont
  {Eisert}(2016)}]{gogolin2016equilibration}%
  \BibitemOpen
  \bibfield  {author} {\bibinfo {author} {\bibfnamefont {C.}~\bibnamefont
  {Gogolin}}\ and\ \bibinfo {author} {\bibfnamefont {J.}~\bibnamefont
  {Eisert}},\ }\bibfield  {title} {\bibinfo {title} {Equilibration,
  thermalisation, and the emergence of statistical mechanics in closed quantum
  systems},\ }\href {https://doi.org/10.1088/0034-4885/79/5/056001} {\bibfield
  {journal} {\bibinfo  {journal} {Rep. Prog. Phys.}\ }\textbf {\bibinfo
  {volume} {79}},\ \bibinfo {pages} {056001} (\bibinfo {year}
  {2016})}\BibitemShut {NoStop}%
\bibitem [{\citenamefont {Rigol}\ \emph {et~al.}(2008)\citenamefont {Rigol},
  \citenamefont {Dunjko},\ and\ \citenamefont
  {Olshanii}}]{rigol2008thermalization}%
  \BibitemOpen
  \bibfield  {author} {\bibinfo {author} {\bibfnamefont {M.}~\bibnamefont
  {Rigol}}, \bibinfo {author} {\bibfnamefont {V.}~\bibnamefont {Dunjko}},\ and\
  \bibinfo {author} {\bibfnamefont {M.}~\bibnamefont {Olshanii}},\ }\bibfield
  {title} {\bibinfo {title} {Thermalization and its mechanism for generic
  isolated quantum systems},\ }\href {https://doi.org/10.1038/nature06838}
  {\bibfield  {journal} {\bibinfo  {journal} {Nature}\ }\textbf {\bibinfo
  {volume} {452}},\ \bibinfo {pages} {854} (\bibinfo {year}
  {2008})}\BibitemShut {NoStop}%
\bibitem [{\citenamefont {Serbyn}\ \emph {et~al.}(2021)\citenamefont {Serbyn},
  \citenamefont {Abanin},\ and\ \citenamefont {Papi{\'c}}}]{serbyn2021quantum}%
  \BibitemOpen
  \bibfield  {author} {\bibinfo {author} {\bibfnamefont {M.}~\bibnamefont
  {Serbyn}}, \bibinfo {author} {\bibfnamefont {D.~A.}\ \bibnamefont {Abanin}},\
  and\ \bibinfo {author} {\bibfnamefont {Z.}~\bibnamefont {Papi{\'c}}},\
  }\bibfield  {title} {\bibinfo {title} {Quantum many-body scars and weak
  breaking of ergodicity},\ }\href {https://doi.org/10.1038/s41567-021-01230-2}
  {\bibfield  {journal} {\bibinfo  {journal} {Nat. Phys.}\ }\textbf {\bibinfo
  {volume} {17}},\ \bibinfo {pages} {675} (\bibinfo {year} {2021})}\BibitemShut
  {NoStop}%
\bibitem [{\citenamefont {Calabrese}\ and\ \citenamefont
  {Cardy}(2016)}]{calabrese2016}%
  \BibitemOpen
  \bibfield  {author} {\bibinfo {author} {\bibfnamefont {P.}~\bibnamefont
  {Calabrese}}\ and\ \bibinfo {author} {\bibfnamefont {J.}~\bibnamefont
  {Cardy}},\ }\bibfield  {title} {\bibinfo {title} {Quantum quenches in $1+1$
  dimensional conformal field theories},\ }\href
  {https://doi.org/10.1088/1742-5468/2016/06/064003} {\bibfield  {journal}
  {\bibinfo  {journal} {Journal of Statistical Mechanics: Theory and
  Experiment}\ }\textbf {\bibinfo {volume} {2016}},\ \bibinfo {pages} {064003}
  (\bibinfo {year} {2016})}\BibitemShut {NoStop}%
\bibitem [{\citenamefont {Bertini}\ \emph {et~al.}(2021)\citenamefont
  {Bertini}, \citenamefont {Heidrich-Meisner}, \citenamefont {Karrasch},
  \citenamefont {Prosen}, \citenamefont {Steinigeweg},\ and\ \citenamefont
  {{\v{Z}}nidari{\v{c}}}}]{bertini2021finitetemperature}%
  \BibitemOpen
  \bibfield  {author} {\bibinfo {author} {\bibfnamefont {B.}~\bibnamefont
  {Bertini}}, \bibinfo {author} {\bibfnamefont {F.}~\bibnamefont
  {Heidrich-Meisner}}, \bibinfo {author} {\bibfnamefont {C.}~\bibnamefont
  {Karrasch}}, \bibinfo {author} {\bibfnamefont {T.}~\bibnamefont {Prosen}},
  \bibinfo {author} {\bibfnamefont {R.}~\bibnamefont {Steinigeweg}},\ and\
  \bibinfo {author} {\bibfnamefont {M.}~\bibnamefont {{\v{Z}}nidari{\v{c}}}},\
  }\bibfield  {title} {\bibinfo {title} {Finite-temperature transport in
  one-dimensional quantum lattice models},\ }\href
  {https://doi.org/10.1103/RevModPhys.93.025003} {\bibfield  {journal}
  {\bibinfo  {journal} {Rev. Mod. Phys.}\ }\textbf {\bibinfo {volume} {93}},\
  \bibinfo {pages} {025003} (\bibinfo {year} {2021})}\BibitemShut {NoStop}%
\bibitem [{\citenamefont {Klobas}\ \emph {et~al.}(2021)\citenamefont {Klobas},
  \citenamefont {Bertini},\ and\ \citenamefont {Piroli}}]{klobas2021exact}%
  \BibitemOpen
  \bibfield  {author} {\bibinfo {author} {\bibfnamefont {K.}~\bibnamefont
  {Klobas}}, \bibinfo {author} {\bibfnamefont {B.}~\bibnamefont {Bertini}},\
  and\ \bibinfo {author} {\bibfnamefont {L.}~\bibnamefont {Piroli}},\
  }\bibfield  {title} {\bibinfo {title} {Exact thermalization dynamics in the
  ``{R}ule 54'' quantum cellular automaton},\ }\href
  {https://doi.org/10.1103/PhysRevLett.126.160602} {\bibfield  {journal}
  {\bibinfo  {journal} {Phys. Rev. Lett.}\ }\textbf {\bibinfo {volume} {126}},\
  \bibinfo {pages} {160602} (\bibinfo {year} {2021})}\BibitemShut {NoStop}%
\bibitem [{\citenamefont {Klobas}\ and\ \citenamefont
  {Bertini}(2021{\natexlab{a}})}]{klobas2021entanglement}%
  \BibitemOpen
  \bibfield  {author} {\bibinfo {author} {\bibfnamefont {K.}~\bibnamefont
  {Klobas}}\ and\ \bibinfo {author} {\bibfnamefont {B.}~\bibnamefont
  {Bertini}},\ }\bibfield  {title} {\bibinfo {title} {Entanglement dynamics in
  {R}ule 54: Exact results and quasiparticle picture},\ }\href
  {https://doi.org/10.21468/SciPostPhys.11.6.107} {\bibfield  {journal}
  {\bibinfo  {journal} {SciPost Phys.}\ }\textbf {\bibinfo {volume} {11}},\
  \bibinfo {pages} {107} (\bibinfo {year} {2021}{\natexlab{a}})}\BibitemShut
  {NoStop}%
\bibitem [{\citenamefont {Klobas}\ and\ \citenamefont
  {Bertini}(2021{\natexlab{b}})}]{klobas2021exactrelaxation}%
  \BibitemOpen
  \bibfield  {author} {\bibinfo {author} {\bibfnamefont {K.}~\bibnamefont
  {Klobas}}\ and\ \bibinfo {author} {\bibfnamefont {B.}~\bibnamefont
  {Bertini}},\ }\bibfield  {title} {\bibinfo {title} {Exact relaxation to
  {Gibbs} and non-equilibrium steady states in the quantum cellular automaton
  {Rule} 54},\ }\href {https://doi.org/10.21468/SciPostPhys.11.6.106}
  {\bibfield  {journal} {\bibinfo  {journal} {SciPost Phys.}\ }\textbf
  {\bibinfo {volume} {11}},\ \bibinfo {pages} {106} (\bibinfo {year}
  {2021}{\natexlab{b}})}\BibitemShut {NoStop}%
\bibitem [{\citenamefont {Calabrese}\ and\ \citenamefont
  {Cardy}(2005)}]{quasi2005}%
  \BibitemOpen
  \bibfield  {author} {\bibinfo {author} {\bibfnamefont {P.}~\bibnamefont
  {Calabrese}}\ and\ \bibinfo {author} {\bibfnamefont {J.}~\bibnamefont
  {Cardy}},\ }\bibfield  {title} {\bibinfo {title} {Evolution of entanglement
  entropy in one-dimensional systems},\ }\href
  {https://doi.org/10.1088/1742-5468/2005/04/p04010} {\bibfield  {journal}
  {\bibinfo  {journal} {Journal of Statistical Mechanics: Theory and
  Experiment}\ }\textbf {\bibinfo {volume} {2005}},\ \bibinfo {pages} {P04010}
  (\bibinfo {year} {2005})}\BibitemShut {NoStop}%
\bibitem [{\citenamefont {Calabrese}\ \emph {et~al.}(2009)\citenamefont
  {Calabrese}, \citenamefont {Cardy},\ and\ \citenamefont
  {Doyon}}]{calabrese2009entanglement}%
  \BibitemOpen
  \bibfield  {author} {\bibinfo {author} {\bibfnamefont {P.}~\bibnamefont
  {Calabrese}}, \bibinfo {author} {\bibfnamefont {J.}~\bibnamefont {Cardy}},\
  and\ \bibinfo {author} {\bibfnamefont {B.}~\bibnamefont {Doyon}},\ }\bibfield
   {title} {\bibinfo {title} {Entanglement entropy in extended quantum
  systems},\ }\href {https://doi.org/10.1088/1751-8121/42/50/500301} {\bibfield
   {journal} {\bibinfo  {journal} {Journal of Physics A: Mathematical and
  Theoretical}\ }\textbf {\bibinfo {volume} {42}},\ \bibinfo {pages} {500301}
  (\bibinfo {year} {2009})}\BibitemShut {NoStop}%
\bibitem [{\citenamefont {Bertini}\ \emph
  {et~al.}(2019{\natexlab{a}})\citenamefont {Bertini}, \citenamefont {Kos},\
  and\ \citenamefont {Prosen}}]{bertini2019exact}%
  \BibitemOpen
  \bibfield  {author} {\bibinfo {author} {\bibfnamefont {B.}~\bibnamefont
  {Bertini}}, \bibinfo {author} {\bibfnamefont {P.}~\bibnamefont {Kos}},\ and\
  \bibinfo {author} {\bibfnamefont {T.}~\bibnamefont {Prosen}},\ }\bibfield
  {title} {\bibinfo {title} {Exact correlation functions for dual-unitary
  lattice models in $1+1$ dimensions},\ }\href
  {https://doi.org/10.1103/PhysRevLett.123.210601} {\bibfield  {journal}
  {\bibinfo  {journal} {Phys. Rev. Lett.}\ }\textbf {\bibinfo {volume} {123}},\
  \bibinfo {pages} {210601} (\bibinfo {year} {2019}{\natexlab{a}})}\BibitemShut
  {NoStop}%
\bibitem [{\citenamefont {Bertini}\ \emph
  {et~al.}(2019{\natexlab{b}})\citenamefont {Bertini}, \citenamefont {Kos},\
  and\ \citenamefont {Prosen}}]{bertini2019entanglement}%
  \BibitemOpen
  \bibfield  {author} {\bibinfo {author} {\bibfnamefont {B.}~\bibnamefont
  {Bertini}}, \bibinfo {author} {\bibfnamefont {P.}~\bibnamefont {Kos}},\ and\
  \bibinfo {author} {\bibfnamefont {T.}~\bibnamefont {Prosen}},\ }\bibfield
  {title} {\bibinfo {title} {Entanglement spreading in a minimal model of
  maximal many-body quantum chaos},\ }\href
  {https://doi.org/10.1103/PhysRevX.9.021033} {\bibfield  {journal} {\bibinfo
  {journal} {Phys. Rev. X}\ }\textbf {\bibinfo {volume} {9}},\ \bibinfo {pages}
  {021033} (\bibinfo {year} {2019}{\natexlab{b}})}\BibitemShut {NoStop}%
\bibitem [{\citenamefont {Piroli}\ \emph {et~al.}(2020)\citenamefont {Piroli},
  \citenamefont {Bertini}, \citenamefont {Cirac},\ and\ \citenamefont
  {Prosen}}]{piroli2020exact}%
  \BibitemOpen
  \bibfield  {author} {\bibinfo {author} {\bibfnamefont {L.}~\bibnamefont
  {Piroli}}, \bibinfo {author} {\bibfnamefont {B.}~\bibnamefont {Bertini}},
  \bibinfo {author} {\bibfnamefont {J.~I.}\ \bibnamefont {Cirac}},\ and\
  \bibinfo {author} {\bibfnamefont {T.}~\bibnamefont {Prosen}},\ }\bibfield
  {title} {\bibinfo {title} {Exact dynamics in dual-unitary quantum circuits},\
  }\href {https://doi.org/10.1103/PhysRevB.101.094304} {\bibfield  {journal}
  {\bibinfo  {journal} {Phys. Rev. B}\ }\textbf {\bibinfo {volume} {101}},\
  \bibinfo {pages} {094304} (\bibinfo {year} {2020})}\BibitemShut {NoStop}%
\bibitem [{\citenamefont {Nahum}\ \emph {et~al.}(2017)\citenamefont {Nahum},
  \citenamefont {Ruhman}, \citenamefont {Vijay},\ and\ \citenamefont
  {Haah}}]{nahum2017quantum}%
  \BibitemOpen
  \bibfield  {author} {\bibinfo {author} {\bibfnamefont {A.}~\bibnamefont
  {Nahum}}, \bibinfo {author} {\bibfnamefont {J.}~\bibnamefont {Ruhman}},
  \bibinfo {author} {\bibfnamefont {S.}~\bibnamefont {Vijay}},\ and\ \bibinfo
  {author} {\bibfnamefont {J.}~\bibnamefont {Haah}},\ }\bibfield  {title}
  {\bibinfo {title} {Quantum entanglement growth under random unitary
  dynamics},\ }\href {https://doi.org/10.1103/PhysRevX.7.031016} {\bibfield
  {journal} {\bibinfo  {journal} {Phys. Rev. X}\ }\textbf {\bibinfo {volume}
  {7}},\ \bibinfo {pages} {031016} (\bibinfo {year} {2017})}\BibitemShut
  {NoStop}%
\bibitem [{\citenamefont {Nahum}\ \emph {et~al.}(2018)\citenamefont {Nahum},
  \citenamefont {Vijay},\ and\ \citenamefont {Haah}}]{nahum2018operator}%
  \BibitemOpen
  \bibfield  {author} {\bibinfo {author} {\bibfnamefont {A.}~\bibnamefont
  {Nahum}}, \bibinfo {author} {\bibfnamefont {S.}~\bibnamefont {Vijay}},\ and\
  \bibinfo {author} {\bibfnamefont {J.}~\bibnamefont {Haah}},\ }\bibfield
  {title} {\bibinfo {title} {Operator spreading in random unitary circuits},\
  }\href {https://doi.org/10.1103/PhysRevX.8.021014} {\bibfield  {journal}
  {\bibinfo  {journal} {Phys. Rev. X}\ }\textbf {\bibinfo {volume} {8}},\
  \bibinfo {pages} {021014} (\bibinfo {year} {2018})}\BibitemShut {NoStop}%
\bibitem [{\citenamefont {von Keyserlingk}\ \emph {et~al.}(2018)\citenamefont
  {von Keyserlingk}, \citenamefont {Rakovszky}, \citenamefont {Pollmann},\ and\
  \citenamefont {Sondhi}}]{vonKeyserlingk2018operator}%
  \BibitemOpen
  \bibfield  {author} {\bibinfo {author} {\bibfnamefont {C.~W.}\ \bibnamefont
  {von Keyserlingk}}, \bibinfo {author} {\bibfnamefont {T.}~\bibnamefont
  {Rakovszky}}, \bibinfo {author} {\bibfnamefont {F.}~\bibnamefont
  {Pollmann}},\ and\ \bibinfo {author} {\bibfnamefont {S.~L.}\ \bibnamefont
  {Sondhi}},\ }\bibfield  {title} {\bibinfo {title} {Operator hydrodynamics,
  {OTOCs}, and entanglement growth in systems without conservation laws},\
  }\href {https://doi.org/10.1103/PhysRevX.8.021013} {\bibfield  {journal}
  {\bibinfo  {journal} {Phys. Rev. X}\ }\textbf {\bibinfo {volume} {8}},\
  \bibinfo {pages} {021013} (\bibinfo {year} {2018})}\BibitemShut {NoStop}%
\bibitem [{\citenamefont {Zhou}\ and\ \citenamefont
  {Nahum}(2019)}]{zhou2019emergent}%
  \BibitemOpen
  \bibfield  {author} {\bibinfo {author} {\bibfnamefont {T.}~\bibnamefont
  {Zhou}}\ and\ \bibinfo {author} {\bibfnamefont {A.}~\bibnamefont {Nahum}},\
  }\bibfield  {title} {\bibinfo {title} {Emergent statistical mechanics of
  entanglement in random unitary circuits},\ }\href
  {https://doi.org/10.1103/PhysRevB.99.174205} {\bibfield  {journal} {\bibinfo
  {journal} {Phys. Rev. B}\ }\textbf {\bibinfo {volume} {99}},\ \bibinfo
  {pages} {174205} (\bibinfo {year} {2019})}\BibitemShut {NoStop}%
\bibitem [{\citenamefont {Fisher}\ \emph {et~al.}(2022)\citenamefont {Fisher},
  \citenamefont {Khemani}, \citenamefont {Nahum},\ and\ \citenamefont
  {Vijay}}]{fisher2022random}%
  \BibitemOpen
  \bibfield  {author} {\bibinfo {author} {\bibfnamefont {M.~P.~A.}\
  \bibnamefont {Fisher}}, \bibinfo {author} {\bibfnamefont {V.}~\bibnamefont
  {Khemani}}, \bibinfo {author} {\bibfnamefont {A.}~\bibnamefont {Nahum}},\
  and\ \bibinfo {author} {\bibfnamefont {S.}~\bibnamefont {Vijay}},\ }\href
  {https://doi.org/10.48550/arXiv.2207.14280} {\bibinfo {title} {Random
  {{Quantum Circuits}}}} (\bibinfo {year} {2022}),\ \Eprint
  {https://arxiv.org/abs/2207.14280} {arXiv:2207.14280 [cond-mat,
  physics:quant-ph]} \BibitemShut {NoStop}%
\bibitem [{\citenamefont {Schollw\"ock}(2011)}]{schollwoeck2011the}%
  \BibitemOpen
  \bibfield  {author} {\bibinfo {author} {\bibfnamefont {U.}~\bibnamefont
  {Schollw\"ock}},\ }\bibfield  {title} {\bibinfo {title} {The density-matrix
  renormalization group in the age of matrix product states},\ }\href
  {https://doi.org/https://doi.org/10.1016/j.aop.2010.09.012} {\bibfield
  {journal} {\bibinfo  {journal} {Annals of Physics}\ }\textbf {\bibinfo
  {volume} {326}},\ \bibinfo {pages} {96} (\bibinfo {year} {2011})}\BibitemShut
  {NoStop}%
\bibitem [{\citenamefont {Daley}\ \emph {et~al.}(2004)\citenamefont {Daley},
  \citenamefont {Kollath}, \citenamefont {Schollw\"ock},\ and\ \citenamefont
  {Vidal}}]{daley2004time}%
  \BibitemOpen
  \bibfield  {author} {\bibinfo {author} {\bibfnamefont {A.~J.}\ \bibnamefont
  {Daley}}, \bibinfo {author} {\bibfnamefont {C.}~\bibnamefont {Kollath}},
  \bibinfo {author} {\bibfnamefont {U.}~\bibnamefont {Schollw\"ock}},\ and\
  \bibinfo {author} {\bibfnamefont {G.}~\bibnamefont {Vidal}},\ }\bibfield
  {title} {\bibinfo {title} {Time-dependent density-matrix
  renormalization-group using adaptive effective hilbert spaces},\ }\href
  {https://doi.org/10.1088/1742-5468/2004/04/P04005} {\bibfield  {journal}
  {\bibinfo  {journal} {Journal of Statistical Mechanics: Theory and
  Experiment}\ }\textbf {\bibinfo {volume} {2004}},\ \bibinfo {pages} {P04005}
  (\bibinfo {year} {2004})}\BibitemShut {NoStop}%
\bibitem [{\citenamefont {White}\ and\ \citenamefont
  {Feiguin}(2004)}]{white2004real}%
  \BibitemOpen
  \bibfield  {author} {\bibinfo {author} {\bibfnamefont {S.~R.}\ \bibnamefont
  {White}}\ and\ \bibinfo {author} {\bibfnamefont {A.~E.}\ \bibnamefont
  {Feiguin}},\ }\bibfield  {title} {\bibinfo {title} {Real-time evolution using
  the density matrix renormalization group},\ }\href
  {https://doi.org/10.1103/PhysRevLett.93.076401} {\bibfield  {journal}
  {\bibinfo  {journal} {Phys. Rev. Lett.}\ }\textbf {\bibinfo {volume} {93}},\
  \bibinfo {pages} {076401} (\bibinfo {year} {2004})}\BibitemShut {NoStop}%
\bibitem [{\citenamefont {Vidal}(2003)}]{vidal2003efficient}%
  \BibitemOpen
  \bibfield  {author} {\bibinfo {author} {\bibfnamefont {G.}~\bibnamefont
  {Vidal}},\ }\bibfield  {title} {\bibinfo {title} {Efficient classical
  simulation of slightly entangled quantum computations},\ }\href
  {https://doi.org/10.1103/PhysRevLett.91.147902} {\bibfield  {journal}
  {\bibinfo  {journal} {Phys. Rev. Lett.}\ }\textbf {\bibinfo {volume} {91}},\
  \bibinfo {pages} {147902} (\bibinfo {year} {2003})}\BibitemShut {NoStop}%
\bibitem [{\citenamefont {Vidal}(2004)}]{vidal2004efficient}%
  \BibitemOpen
  \bibfield  {author} {\bibinfo {author} {\bibfnamefont {G.}~\bibnamefont
  {Vidal}},\ }\bibfield  {title} {\bibinfo {title} {Efficient simulation of
  one-dimensional quantum many-body systems},\ }\href
  {https://doi.org/10.1103/PhysRevLett.93.040502} {\bibfield  {journal}
  {\bibinfo  {journal} {Phys. Rev. Lett.}\ }\textbf {\bibinfo {volume} {93}},\
  \bibinfo {pages} {040502} (\bibinfo {year} {2004})}\BibitemShut {NoStop}%
\bibitem [{\citenamefont {Amico}\ \emph {et~al.}(2008)\citenamefont {Amico},
  \citenamefont {Fazio}, \citenamefont {Osterloh},\ and\ \citenamefont
  {Vedral}}]{amico2008entanglement}%
  \BibitemOpen
  \bibfield  {author} {\bibinfo {author} {\bibfnamefont {L.}~\bibnamefont
  {Amico}}, \bibinfo {author} {\bibfnamefont {R.}~\bibnamefont {Fazio}},
  \bibinfo {author} {\bibfnamefont {A.}~\bibnamefont {Osterloh}},\ and\
  \bibinfo {author} {\bibfnamefont {V.}~\bibnamefont {Vedral}},\ }\bibfield
  {title} {\bibinfo {title} {Entanglement in many-body systems},\ }\href
  {https://doi.org/10.1103/RevModPhys.80.517} {\bibfield  {journal} {\bibinfo
  {journal} {Rev. Mod. Phys.}\ }\textbf {\bibinfo {volume} {80}},\ \bibinfo
  {pages} {517} (\bibinfo {year} {2008})}\BibitemShut {NoStop}%
\bibitem [{\citenamefont {Laflorencie}(2016)}]{laflorencie2016quantum}%
  \BibitemOpen
  \bibfield  {author} {\bibinfo {author} {\bibfnamefont {N.}~\bibnamefont
  {Laflorencie}},\ }\bibfield  {title} {\bibinfo {title} {Quantum entanglement
  in condensed matter systems},\ }\href
  {https://doi.org/https://doi.org/10.1016/j.physrep.2016.06.008} {\bibfield
  {journal} {\bibinfo  {journal} {Physics Reports}\ }\textbf {\bibinfo {volume}
  {646}},\ \bibinfo {pages} {1} (\bibinfo {year} {2016})},\ \bibinfo {note}
  {quantum entanglement in condensed matter systems}\BibitemShut {NoStop}%
\bibitem [{\citenamefont {Calabrese}(2018)}]{calabrese2018entanglement}%
  \BibitemOpen
  \bibfield  {author} {\bibinfo {author} {\bibfnamefont {P.}~\bibnamefont
  {Calabrese}},\ }\bibfield  {title} {\bibinfo {title} {Entanglement and
  thermodynamics in non-equilibrium isolated quantum systems},\ }\href
  {https://doi.org/https://doi.org/10.1016/j.physa.2017.10.011} {\bibfield
  {journal} {\bibinfo  {journal} {Physica A: Statistical Mechanics and its
  Applications}\ }\textbf {\bibinfo {volume} {504}},\ \bibinfo {pages} {31}
  (\bibinfo {year} {2018})},\ \bibinfo {note} {lecture Notes of the 14th
  International Summer School on Fundamental Problems in Statistical
  Physics}\BibitemShut {NoStop}%
\bibitem [{\citenamefont {Santos}\ \emph {et~al.}(2011)\citenamefont {Santos},
  \citenamefont {Polkovnikov},\ and\ \citenamefont
  {Rigol}}]{santos2011entropy}%
  \BibitemOpen
  \bibfield  {author} {\bibinfo {author} {\bibfnamefont {L.~F.}\ \bibnamefont
  {Santos}}, \bibinfo {author} {\bibfnamefont {A.}~\bibnamefont
  {Polkovnikov}},\ and\ \bibinfo {author} {\bibfnamefont {M.}~\bibnamefont
  {Rigol}},\ }\bibfield  {title} {\bibinfo {title} {Entropy of isolated quantum
  systems after a quench},\ }\href
  {https://doi.org/10.1103/PhysRevLett.107.040601} {\bibfield  {journal}
  {\bibinfo  {journal} {Phys. Rev. Lett.}\ }\textbf {\bibinfo {volume} {107}},\
  \bibinfo {pages} {040601} (\bibinfo {year} {2011})}\BibitemShut {NoStop}%
\bibitem [{\citenamefont {Gurarie}(2013)}]{gurarie2013global}%
  \BibitemOpen
  \bibfield  {author} {\bibinfo {author} {\bibfnamefont {V.}~\bibnamefont
  {Gurarie}},\ }\bibfield  {title} {\bibinfo {title} {Global large time
  dynamics and the generalized gibbs ensemble},\ }\href
  {https://doi.org/10.1088/1742-5468/2013/02/P02014} {\bibfield  {journal}
  {\bibinfo  {journal} {Journal of Statistical Mechanics: Theory and
  Experiment}\ }\textbf {\bibinfo {volume} {2013}},\ \bibinfo {pages} {P02014}
  (\bibinfo {year} {2013})}\BibitemShut {NoStop}%
\bibitem [{\citenamefont {Bertini}\ \emph {et~al.}(2022)\citenamefont
  {Bertini}, \citenamefont {Klobas}, \citenamefont {Alba}, \citenamefont
  {Lagnese},\ and\ \citenamefont {Calabrese}}]{bertini2022growth}%
  \BibitemOpen
  \bibfield  {author} {\bibinfo {author} {\bibfnamefont {B.}~\bibnamefont
  {Bertini}}, \bibinfo {author} {\bibfnamefont {K.}~\bibnamefont {Klobas}},
  \bibinfo {author} {\bibfnamefont {V.}~\bibnamefont {Alba}}, \bibinfo {author}
  {\bibfnamefont {G.}~\bibnamefont {Lagnese}},\ and\ \bibinfo {author}
  {\bibfnamefont {P.}~\bibnamefont {Calabrese}},\ }\bibfield  {title} {\bibinfo
  {title} {Growth of {R\'enyi} entropies in interacting integrable models and
  the breakdown of the quasiparticle picture},\ }\href
  {https://doi.org/10.1103/PhysRevX.12.031016} {\bibfield  {journal} {\bibinfo
  {journal} {Phys. Rev. X}\ }\textbf {\bibinfo {volume} {12}},\ \bibinfo
  {pages} {031016} (\bibinfo {year} {2022})}\BibitemShut {NoStop}%
\bibitem [{\citenamefont {Bertini}\ \emph
  {et~al.}(2023{\natexlab{a}})\citenamefont {Bertini}, \citenamefont
  {Calabrese}, \citenamefont {Collura}, \citenamefont {Klobas},\ and\
  \citenamefont {Rylands}}]{bertini2022nonequilibrium}%
  \BibitemOpen
  \bibfield  {author} {\bibinfo {author} {\bibfnamefont {B.}~\bibnamefont
  {Bertini}}, \bibinfo {author} {\bibfnamefont {P.}~\bibnamefont {Calabrese}},
  \bibinfo {author} {\bibfnamefont {M.}~\bibnamefont {Collura}}, \bibinfo
  {author} {\bibfnamefont {K.}~\bibnamefont {Klobas}},\ and\ \bibinfo {author}
  {\bibfnamefont {C.}~\bibnamefont {Rylands}},\ }\bibfield  {title} {\bibinfo
  {title} {Nonequilibrium full counting statistics and symmetry-resolved
  entanglement from space-time duality},\ }\href
  {https://doi.org/10.1103/PhysRevLett.131.140401} {\bibfield  {journal}
  {\bibinfo  {journal} {Phys. Rev. Lett.}\ }\textbf {\bibinfo {volume} {131}},\
  \bibinfo {pages} {140401} (\bibinfo {year} {2023}{\natexlab{a}})}\BibitemShut
  {NoStop}%
\bibitem [{\citenamefont {Bertini}\ \emph
  {et~al.}(2023{\natexlab{b}})\citenamefont {Bertini}, \citenamefont {Klobas},
  \citenamefont {Collura}, \citenamefont {Calabrese},\ and\ \citenamefont
  {Rylands}}]{bertini2023dynamics}%
  \BibitemOpen
  \bibfield  {author} {\bibinfo {author} {\bibfnamefont {B.}~\bibnamefont
  {Bertini}}, \bibinfo {author} {\bibfnamefont {K.}~\bibnamefont {Klobas}},
  \bibinfo {author} {\bibfnamefont {M.}~\bibnamefont {Collura}}, \bibinfo
  {author} {\bibfnamefont {P.}~\bibnamefont {Calabrese}},\ and\ \bibinfo
  {author} {\bibfnamefont {C.}~\bibnamefont {Rylands}},\ }\bibfield  {title}
  {\bibinfo {title} {Dynamics of charge fluctuations from asymmetric initial
  states},\ }\Eprint {https://arxiv.org/abs/2306.12404} {arXiv:2306.12404}
  (\bibinfo {year} {2023}{\natexlab{b}})\BibitemShut {NoStop}%
\bibitem [{\citenamefont {Alba}\ and\ \citenamefont
  {Calabrese}(2017)}]{alba2017entanglement}%
  \BibitemOpen
  \bibfield  {author} {\bibinfo {author} {\bibfnamefont {V.}~\bibnamefont
  {Alba}}\ and\ \bibinfo {author} {\bibfnamefont {P.}~\bibnamefont
  {Calabrese}},\ }\bibfield  {title} {\bibinfo {title} {Entanglement and
  thermodynamics after a quantum quench in integrable systems},\ }\href
  {https://doi.org/10.1073/pnas.1703516114} {\bibfield  {journal} {\bibinfo
  {journal} {Proc. Natl. Acad. Sci. U.S.A.}\ }\textbf {\bibinfo {volume}
  {114}},\ \bibinfo {pages} {7947} (\bibinfo {year} {2017})}\BibitemShut
  {NoStop}%
\bibitem [{\citenamefont {Fagotti}\ and\ \citenamefont
  {Calabrese}(2008)}]{fagotti2008evolution}%
  \BibitemOpen
  \bibfield  {author} {\bibinfo {author} {\bibfnamefont {M.}~\bibnamefont
  {Fagotti}}\ and\ \bibinfo {author} {\bibfnamefont {P.}~\bibnamefont
  {Calabrese}},\ }\bibfield  {title} {\bibinfo {title} {Evolution of
  entanglement entropy following a quantum quench: Analytic results for the
  {XY} chain in a transverse magnetic field},\ }\href
  {https://doi.org/10.1103/PhysRevA.78.010306} {\bibfield  {journal} {\bibinfo
  {journal} {Phys. Rev. A}\ }\textbf {\bibinfo {volume} {78}},\ \bibinfo
  {pages} {010306} (\bibinfo {year} {2008})}\BibitemShut {NoStop}%
\bibitem [{\citenamefont {Castro-Alvaredo}\ \emph {et~al.}(2019)\citenamefont
  {Castro-Alvaredo}, \citenamefont {Lencs{\'e}s}, \citenamefont
  {Sz{\'e}cs{\'e}nyi},\ and\ \citenamefont
  {Viti}}]{castroalvaredo2019entanglement}%
  \BibitemOpen
  \bibfield  {author} {\bibinfo {author} {\bibfnamefont {O.~A.}\ \bibnamefont
  {Castro-Alvaredo}}, \bibinfo {author} {\bibfnamefont {M.}~\bibnamefont
  {Lencs{\'e}s}}, \bibinfo {author} {\bibfnamefont {I.~M.}\ \bibnamefont
  {Sz{\'e}cs{\'e}nyi}},\ and\ \bibinfo {author} {\bibfnamefont
  {J.}~\bibnamefont {Viti}},\ }\bibfield  {title} {\bibinfo {title}
  {Entanglement dynamics after a quench in {Ising} field theory: a branch point
  twist field approach},\ }\href {https://doi.org/10.1007/JHEP12(2019)079}
  {\bibfield  {journal} {\bibinfo  {journal} {J. High Energy Phys.}\ }\textbf
  {\bibinfo {volume} {2019}}\bibinfo  {number} { (12)},\ \bibinfo {pages}
  {1}}\BibitemShut {NoStop}%
\bibitem [{\citenamefont {Alba}\ and\ \citenamefont
  {Calabrese}(2018)}]{alba2018entanglement}%
  \BibitemOpen
\bibfield  {number} {  }\bibfield  {author} {\bibinfo {author} {\bibfnamefont
  {V.}~\bibnamefont {Alba}}\ and\ \bibinfo {author} {\bibfnamefont
  {P.}~\bibnamefont {Calabrese}},\ }\bibfield  {title} {\bibinfo {title}
  {Entanglement dynamics after quantum quenches in generic integrable
  systems},\ }\href {https://doi.org/10.21468/SciPostPhys.4.3.017} {\bibfield
  {journal} {\bibinfo  {journal} {SciPost Phys.}\ }\textbf {\bibinfo {volume}
  {4}},\ \bibinfo {pages} {17} (\bibinfo {year} {2018})}\BibitemShut {NoStop}%
\bibitem [{\citenamefont {Alba}\ \emph {et~al.}(2019)\citenamefont {Alba},
  \citenamefont {Bertini},\ and\ \citenamefont
  {Fagotti}}]{alba2019entanglement}%
  \BibitemOpen
  \bibfield  {author} {\bibinfo {author} {\bibfnamefont {V.}~\bibnamefont
  {Alba}}, \bibinfo {author} {\bibfnamefont {B.}~\bibnamefont {Bertini}},\ and\
  \bibinfo {author} {\bibfnamefont {M.}~\bibnamefont {Fagotti}},\ }\bibfield
  {title} {\bibinfo {title} {Entanglement evolution and generalised
  hydrodynamics: {I}nteracting integrable systems},\ }\href
  {https://doi.org/10.21468/SciPostPhys.7.1.005} {\bibfield  {journal}
  {\bibinfo  {journal} {SciPost Phys.}\ }\textbf {\bibinfo {volume} {7}},\
  \bibinfo {pages} {5} (\bibinfo {year} {2019})}\BibitemShut {NoStop}%
\bibitem [{\citenamefont {Liu}\ and\ \citenamefont
  {Suh}(2014)}]{liu2014entanglement}%
  \BibitemOpen
  \bibfield  {author} {\bibinfo {author} {\bibfnamefont {H.}~\bibnamefont
  {Liu}}\ and\ \bibinfo {author} {\bibfnamefont {S.~J.}\ \bibnamefont {Suh}},\
  }\bibfield  {title} {\bibinfo {title} {Entanglement tsunami: Universal
  scaling in holographic thermalization},\ }\href
  {https://doi.org/10.1103/PhysRevLett.112.011601} {\bibfield  {journal}
  {\bibinfo  {journal} {Phys. Rev. Lett.}\ }\textbf {\bibinfo {volume} {112}},\
  \bibinfo {pages} {011601} (\bibinfo {year} {2014})}\BibitemShut {NoStop}%
\bibitem [{\citenamefont {Casini}\ \emph {et~al.}(2016)\citenamefont {Casini},
  \citenamefont {Liu},\ and\ \citenamefont {Mezei}}]{casini2016spread}%
  \BibitemOpen
  \bibfield  {author} {\bibinfo {author} {\bibfnamefont {H.}~\bibnamefont
  {Casini}}, \bibinfo {author} {\bibfnamefont {H.}~\bibnamefont {Liu}},\ and\
  \bibinfo {author} {\bibfnamefont {M.}~\bibnamefont {Mezei}},\ }\bibfield
  {title} {\bibinfo {title} {Spread of entanglement and causality},\ }\href
  {https://doi.org/https://doi.org/10.1007/JHEP07(2016)077} {\bibfield
  {journal} {\bibinfo  {journal} {J. High Energ. Phys.}\ }\textbf {\bibinfo
  {volume} {2016}}\bibinfo  {number} { (7)},\ \bibinfo {pages} {1}}\BibitemShut
  {NoStop}%
\bibitem [{\citenamefont {Cotler}\ \emph {et~al.}(2016)\citenamefont {Cotler},
  \citenamefont {Hertzberg}, \citenamefont {Mezei},\ and\ \citenamefont
  {Mueller}}]{cotler2016entanglement}%
  \BibitemOpen
\bibfield  {number} {  }\bibfield  {author} {\bibinfo {author} {\bibfnamefont
  {J.~S.}\ \bibnamefont {Cotler}}, \bibinfo {author} {\bibfnamefont {M.~P.}\
  \bibnamefont {Hertzberg}}, \bibinfo {author} {\bibfnamefont {M.}~\bibnamefont
  {Mezei}},\ and\ \bibinfo {author} {\bibfnamefont {M.~T.}\ \bibnamefont
  {Mueller}},\ }\bibfield  {title} {\bibinfo {title} {Entanglement growth after
  a global quench in free scalar field theory},\ }\href
  {https://doi.org/10.1007/JHEP11(2016)166} {\bibfield  {journal} {\bibinfo
  {journal} {J. High Energ. Phys.}\ }\textbf {\bibinfo {volume} {2016}}\bibinfo
   {number} { (11)},\ \bibinfo {pages} {1}}\BibitemShut {NoStop}%
\bibitem [{\citenamefont {Maraga}\ \emph {et~al.}(2015)\citenamefont {Maraga},
  \citenamefont {Chiocchetta}, \citenamefont {Mitra},\ and\ \citenamefont
  {Gambassi}}]{maraga2015coarsening}%
  \BibitemOpen
\bibfield  {number} {  }\bibfield  {author} {\bibinfo {author} {\bibfnamefont
  {A.}~\bibnamefont {Maraga}}, \bibinfo {author} {\bibfnamefont
  {A.}~\bibnamefont {Chiocchetta}}, \bibinfo {author} {\bibfnamefont
  {A.}~\bibnamefont {Mitra}},\ and\ \bibinfo {author} {\bibfnamefont
  {A.}~\bibnamefont {Gambassi}},\ }\bibfield  {title} {\bibinfo {title} {Aging
  and coarsening in isolated quantum systems after a quench: Exact results for
  the quantum $\text{O}(n)$ model with $n$ $\ensuremath{\rightarrow}$
  $\ensuremath{\infty}$},\ }\href {https://doi.org/10.1103/PhysRevE.92.042151}
  {\bibfield  {journal} {\bibinfo  {journal} {Phys. Rev. E}\ }\textbf {\bibinfo
  {volume} {92}},\ \bibinfo {pages} {042151} (\bibinfo {year}
  {2015})}\BibitemShut {NoStop}%
\bibitem [{\citenamefont {Chiocchetta}\ \emph {et~al.}(2015)\citenamefont
  {Chiocchetta}, \citenamefont {Tavora}, \citenamefont {Gambassi},\ and\
  \citenamefont {Mitra}}]{chiocchetta2015scaling}%
  \BibitemOpen
  \bibfield  {author} {\bibinfo {author} {\bibfnamefont {A.}~\bibnamefont
  {Chiocchetta}}, \bibinfo {author} {\bibfnamefont {M.}~\bibnamefont {Tavora}},
  \bibinfo {author} {\bibfnamefont {A.}~\bibnamefont {Gambassi}},\ and\
  \bibinfo {author} {\bibfnamefont {A.}~\bibnamefont {Mitra}},\ }\bibfield
  {title} {\bibinfo {title} {Short-time universal scaling in an isolated
  quantum system after a quench},\ }\href
  {https://doi.org/10.1103/PhysRevB.91.220302} {\bibfield  {journal} {\bibinfo
  {journal} {Phys. Rev. B}\ }\textbf {\bibinfo {volume} {91}},\ \bibinfo
  {pages} {220302} (\bibinfo {year} {2015})}\BibitemShut {NoStop}%
\bibitem [{\citenamefont {Chiocchetta}\ \emph {et~al.}(2016)\citenamefont
  {Chiocchetta}, \citenamefont {Tavora}, \citenamefont {Gambassi},\ and\
  \citenamefont {Mitra}}]{chiocchetta2016scaling}%
  \BibitemOpen
  \bibfield  {author} {\bibinfo {author} {\bibfnamefont {A.}~\bibnamefont
  {Chiocchetta}}, \bibinfo {author} {\bibfnamefont {M.}~\bibnamefont {Tavora}},
  \bibinfo {author} {\bibfnamefont {A.}~\bibnamefont {Gambassi}},\ and\
  \bibinfo {author} {\bibfnamefont {A.}~\bibnamefont {Mitra}},\ }\bibfield
  {title} {\bibinfo {title} {Short-time universal scaling and light-cone
  dynamics after a quench in an isolated quantum system in $d$ spatial
  dimensions},\ }\href {https://doi.org/10.1103/PhysRevB.94.134311} {\bibfield
  {journal} {\bibinfo  {journal} {Phys. Rev. B}\ }\textbf {\bibinfo {volume}
  {94}},\ \bibinfo {pages} {134311} (\bibinfo {year} {2016})}\BibitemShut
  {NoStop}%
\bibitem [{\citenamefont {Lemonik}\ and\ \citenamefont
  {Mitra}(2016)}]{lemonik2016bosons}%
  \BibitemOpen
  \bibfield  {author} {\bibinfo {author} {\bibfnamefont {Y.}~\bibnamefont
  {Lemonik}}\ and\ \bibinfo {author} {\bibfnamefont {A.}~\bibnamefont
  {Mitra}},\ }\bibfield  {title} {\bibinfo {title} {Entanglement properties of
  the critical quench of $o(n)$ bosons},\ }\href
  {https://doi.org/10.1103/PhysRevB.94.024306} {\bibfield  {journal} {\bibinfo
  {journal} {Phys. Rev. B}\ }\textbf {\bibinfo {volume} {94}},\ \bibinfo
  {pages} {024306} (\bibinfo {year} {2016})}\BibitemShut {NoStop}%
\bibitem [{\citenamefont {Chiocchetta}\ \emph {et~al.}(2017)\citenamefont
  {Chiocchetta}, \citenamefont {Gambassi}, \citenamefont {Diehl},\ and\
  \citenamefont {Marino}}]{chiocchetta2017crossovers}%
  \BibitemOpen
  \bibfield  {author} {\bibinfo {author} {\bibfnamefont {A.}~\bibnamefont
  {Chiocchetta}}, \bibinfo {author} {\bibfnamefont {A.}~\bibnamefont
  {Gambassi}}, \bibinfo {author} {\bibfnamefont {S.}~\bibnamefont {Diehl}},\
  and\ \bibinfo {author} {\bibfnamefont {J.}~\bibnamefont {Marino}},\
  }\bibfield  {title} {\bibinfo {title} {Dynamical crossovers in prethermal
  critical states},\ }\href {https://doi.org/10.1103/PhysRevLett.118.135701}
  {\bibfield  {journal} {\bibinfo  {journal} {Phys. Rev. Lett.}\ }\textbf
  {\bibinfo {volume} {118}},\ \bibinfo {pages} {135701} (\bibinfo {year}
  {2017})}\BibitemShut {NoStop}%
\bibitem [{\citenamefont {Bertini}\ \emph
  {et~al.}(2018{\natexlab{a}})\citenamefont {Bertini}, \citenamefont
  {Tartaglia},\ and\ \citenamefont {Calabrese}}]{pair2018}%
  \BibitemOpen
  \bibfield  {author} {\bibinfo {author} {\bibfnamefont {B.}~\bibnamefont
  {Bertini}}, \bibinfo {author} {\bibfnamefont {E.}~\bibnamefont {Tartaglia}},\
  and\ \bibinfo {author} {\bibfnamefont {P.}~\bibnamefont {Calabrese}},\
  }\bibfield  {title} {\bibinfo {title} {Entanglement and diagonal entropies
  after a quench with no pair structure},\ }\href
  {https://doi.org/10.1088/1742-5468/aac73f} {\bibfield  {journal} {\bibinfo
  {journal} {Journal of Statistical Mechanics: Theory and Experiment}\ }\textbf
  {\bibinfo {volume} {2018}},\ \bibinfo {pages} {063104} (\bibinfo {year}
  {2018}{\natexlab{a}})}\BibitemShut {NoStop}%
\bibitem [{\citenamefont {Bertini}\ \emph
  {et~al.}(2018{\natexlab{b}})\citenamefont {Bertini}, \citenamefont {Fagotti},
  \citenamefont {Piroli},\ and\ \citenamefont {Calabrese}}]{hydro2018}%
  \BibitemOpen
  \bibfield  {author} {\bibinfo {author} {\bibfnamefont {B.}~\bibnamefont
  {Bertini}}, \bibinfo {author} {\bibfnamefont {M.}~\bibnamefont {Fagotti}},
  \bibinfo {author} {\bibfnamefont {L.}~\bibnamefont {Piroli}},\ and\ \bibinfo
  {author} {\bibfnamefont {P.}~\bibnamefont {Calabrese}},\ }\bibfield  {title}
  {\bibinfo {title} {Entanglement evolution and generalised hydrodynamics:
  noninteracting systems},\ }\href {https://doi.org/10.1088/1751-8121/aad82e}
  {\bibfield  {journal} {\bibinfo  {journal} {Journal of Physics A:
  Mathematical and Theoretical}\ }\textbf {\bibinfo {volume} {51}},\ \bibinfo
  {pages} {39LT01} (\bibinfo {year} {2018}{\natexlab{b}})}\BibitemShut
  {NoStop}%
\bibitem [{\citenamefont {Bastianello}\ and\ \citenamefont
  {Calabrese}(2018)}]{bastianello2018spreading}%
  \BibitemOpen
  \bibfield  {author} {\bibinfo {author} {\bibfnamefont {A.}~\bibnamefont
  {Bastianello}}\ and\ \bibinfo {author} {\bibfnamefont {P.}~\bibnamefont
  {Calabrese}},\ }\bibfield  {title} {\bibinfo {title} {{Spreading of
  entanglement and correlations after a quench with intertwined
  quasiparticles}},\ }\href {https://doi.org/10.21468/SciPostPhys.5.4.033}
  {\bibfield  {journal} {\bibinfo  {journal} {SciPost Phys.}\ }\textbf
  {\bibinfo {volume} {5}},\ \bibinfo {pages} {033} (\bibinfo {year}
  {2018})}\BibitemShut {NoStop}%
\bibitem [{\citenamefont {Haque}\ \emph {et~al.}(2009)\citenamefont {Haque},
  \citenamefont {Zozulya},\ and\ \citenamefont
  {Schoutens}}]{haque2009entanglement}%
  \BibitemOpen
  \bibfield  {author} {\bibinfo {author} {\bibfnamefont {M.}~\bibnamefont
  {Haque}}, \bibinfo {author} {\bibfnamefont {O.~S.}\ \bibnamefont {Zozulya}},\
  and\ \bibinfo {author} {\bibfnamefont {K.}~\bibnamefont {Schoutens}},\
  }\bibfield  {title} {\bibinfo {title} {Entanglement between particle
  partitions in itinerant many-particle states},\ }\href
  {https://doi.org/10.1088/1751-8113/42/50/504012} {\bibfield  {journal}
  {\bibinfo  {journal} {Journal of Physics A: Mathematical and Theoretical}\
  }\textbf {\bibinfo {volume} {42}},\ \bibinfo {pages} {504012} (\bibinfo
  {year} {2009})}\BibitemShut {NoStop}%
\bibitem [{\citenamefont {Casini}\ and\ \citenamefont
  {Huerta}(2009)}]{review2009}%
  \BibitemOpen
  \bibfield  {author} {\bibinfo {author} {\bibfnamefont {H.}~\bibnamefont
  {Casini}}\ and\ \bibinfo {author} {\bibfnamefont {M.}~\bibnamefont
  {Huerta}},\ }\bibfield  {title} {\bibinfo {title} {Entanglement entropy in
  free quantum field theory},\ }\href
  {https://doi.org/10.1088/1751-8113/42/50/504007} {\bibfield  {journal}
  {\bibinfo  {journal} {Journal of Physics A: Mathematical and Theoretical}\
  }\textbf {\bibinfo {volume} {42}},\ \bibinfo {pages} {504007} (\bibinfo
  {year} {2009})}\BibitemShut {NoStop}%
\bibitem [{\citenamefont {Korepin}\ \emph {et~al.}(1993)\citenamefont
  {Korepin}, \citenamefont {Bogoliubov},\ and\ \citenamefont
  {Izergin}}]{korepin1997quantum}%
  \BibitemOpen
  \bibfield  {author} {\bibinfo {author} {\bibfnamefont {V.~E.}\ \bibnamefont
  {Korepin}}, \bibinfo {author} {\bibfnamefont {N.~M.}\ \bibnamefont
  {Bogoliubov}},\ and\ \bibinfo {author} {\bibfnamefont {A.~G.}\ \bibnamefont
  {Izergin}},\ }\href {https://doi.org/10.1017/CBO9780511628832} {\emph
  {\bibinfo {title} {Quantum Inverse Scattering Method and Correlation
  Functions}}},\ Cambridge Monographs on Mathematical Physics\ (\bibinfo
  {publisher} {Cambridge University Press},\ \bibinfo {year}
  {1993})\BibitemShut {NoStop}%
\bibitem [{\citenamefont {Takahashi}(1999)}]{takahashi1999thermodynamics}%
  \BibitemOpen
  \bibfield  {author} {\bibinfo {author} {\bibfnamefont {M.}~\bibnamefont
  {Takahashi}},\ }\href {https://doi.org/10.1017/CBO9780511524332} {\emph
  {\bibinfo {title} {Thermodynamics of One-Dimensional Solvable Models}}}\
  (\bibinfo  {publisher} {Cambridge University Press},\ \bibinfo {year}
  {1999})\BibitemShut {NoStop}%
\bibitem [{Note1()}]{Note1}%
  \BibitemOpen
  \bibinfo {note} {In the case of interacting integrable models the
  quasiparticles undergo non-trivial scattering. Their scattering, however, is
  always elastic and its sole effect is to renormalise the quasiparticle
  velocities~\cite {alba2017entanglement}}\BibitemShut {NoStop}%
\bibitem [{Note2()}]{Note2}%
  \BibitemOpen
  \bibinfo {note} {We remark that the presence of points of non-analyticity in
  the quasiparticle prediction is not in contradiction with the fact that, for
  any finite subsystem, the entanglement dynamics is smooth. Indeed, the
  quasiparticle prediction describes the asymptotic limit Eq.~\protect \textup
  {\hbox {\mathsurround \z@ \protect \normalfont (\ignorespaces \ref
  {eq:scaling}\unskip \@@italiccorr )}}.}\BibitemShut {Stop}%
\bibitem [{\citenamefont {Ares}\ \emph {et~al.}(2023)\citenamefont {Ares},
  \citenamefont {Murciano},\ and\ \citenamefont
  {Calabrese}}]{ares2022entanglement}%
  \BibitemOpen
  \bibfield  {author} {\bibinfo {author} {\bibfnamefont {F.}~\bibnamefont
  {Ares}}, \bibinfo {author} {\bibfnamefont {S.}~\bibnamefont {Murciano}},\
  and\ \bibinfo {author} {\bibfnamefont {P.}~\bibnamefont {Calabrese}},\
  }\bibfield  {title} {\bibinfo {title} {Entanglement asymmetry as a probe of
  symmetry breaking},\ }\href {https://doi.org/10.1038/s41467-023-37747-8}
  {\bibfield  {journal} {\bibinfo  {journal} {Nat. Commun.}\ }\textbf {\bibinfo
  {volume} {14}},\ \bibinfo {pages} {2036} (\bibinfo {year}
  {2023})}\BibitemShut {NoStop}%
\bibitem [{\citenamefont {Rylands}\ \emph {et~al.}()\citenamefont {Rylands},
  \citenamefont {Klobas}, \citenamefont {Ares}, \citenamefont {Calabrese},
  \citenamefont {Murciano},\ and\ \citenamefont
  {Bertini}}]{rylands2023microscopic}%
  \BibitemOpen
  \bibfield  {author} {\bibinfo {author} {\bibfnamefont {C.}~\bibnamefont
  {Rylands}}, \bibinfo {author} {\bibfnamefont {K.}~\bibnamefont {Klobas}},
  \bibinfo {author} {\bibfnamefont {F.}~\bibnamefont {Ares}}, \bibinfo {author}
  {\bibfnamefont {P.}~\bibnamefont {Calabrese}}, \bibinfo {author}
  {\bibfnamefont {S.}~\bibnamefont {Murciano}},\ and\ \bibinfo {author}
  {\bibfnamefont {B.}~\bibnamefont {Bertini}},\ }\bibfield  {title} {\bibinfo
  {title} {Microscopic origin of the quantum mpemba effect in integrable
  systems},\ }\Eprint {https://arxiv.org/abs/2310.04419} {arXiv:2310.04419}
  \BibitemShut {NoStop}%
\bibitem [{\citenamefont {Ferro}\ \emph {et~al.}()\citenamefont {Ferro},
  \citenamefont {Ares},\ and\ \citenamefont
  {Calabrese}}]{ferro2023nonequilibrium}%
  \BibitemOpen
  \bibfield  {author} {\bibinfo {author} {\bibfnamefont {F.}~\bibnamefont
  {Ferro}}, \bibinfo {author} {\bibfnamefont {F.}~\bibnamefont {Ares}},\ and\
  \bibinfo {author} {\bibfnamefont {P.}~\bibnamefont {Calabrese}},\ }\bibfield
  {title} {\bibinfo {title} {Non-equilibrium entanglement asymmetry for
  discrete groups: the example of the xy spin chain},\ }\Eprint
  {https://arxiv.org/abs/2307.06902} {arXiv:2307.06902} \BibitemShut {NoStop}%
\bibitem [{\citenamefont {Yamashika}\ \emph {et~al.}()\citenamefont
  {Yamashika}, \citenamefont {Ares},\ and\ \citenamefont
  {Calabrese}}]{yamashika2023time}%
  \BibitemOpen
  \bibfield  {author} {\bibinfo {author} {\bibfnamefont {S.}~\bibnamefont
  {Yamashika}}, \bibinfo {author} {\bibfnamefont {F.}~\bibnamefont {Ares}},\
  and\ \bibinfo {author} {\bibfnamefont {P.}~\bibnamefont {Calabrese}},\
  }\bibfield  {title} {\bibinfo {title} {Time evolution of entanglement entropy
  after quenches in two-dimensional free fermion systems: a dimensional
  reduction treatment},\ }\Eprint {https://arxiv.org/abs/2310.XXXX}
  {arXiv:2310.XXXX} \BibitemShut {NoStop}%
\bibitem [{\citenamefont {Chung}\ and\ \citenamefont
  {Peschel}(2000)}]{chung2000density}%
  \BibitemOpen
  \bibfield  {author} {\bibinfo {author} {\bibfnamefont {M.-C.}\ \bibnamefont
  {Chung}}\ and\ \bibinfo {author} {\bibfnamefont {I.}~\bibnamefont
  {Peschel}},\ }\bibfield  {title} {\bibinfo {title} {Density-matrix spectra
  for two-dimensional quantum systems},\ }\href
  {https://doi.org/10.1103/PhysRevB.62.4191} {\bibfield  {journal} {\bibinfo
  {journal} {Phys. Rev. B}\ }\textbf {\bibinfo {volume} {62}},\ \bibinfo
  {pages} {4191} (\bibinfo {year} {2000})}\BibitemShut {NoStop}%
\bibitem [{\citenamefont {Ares}\ \emph {et~al.}(2014)\citenamefont {Ares},
  \citenamefont {Esteve}, \citenamefont {Falceto},\ and\ \citenamefont
  {Sánchez-Burillo}}]{ares2014excited}%
  \BibitemOpen
  \bibfield  {author} {\bibinfo {author} {\bibfnamefont {F.}~\bibnamefont
  {Ares}}, \bibinfo {author} {\bibfnamefont {J.~G.}\ \bibnamefont {Esteve}},
  \bibinfo {author} {\bibfnamefont {F.}~\bibnamefont {Falceto}},\ and\ \bibinfo
  {author} {\bibfnamefont {E.}~\bibnamefont {Sánchez-Burillo}},\ }\bibfield
  {title} {\bibinfo {title} {Excited state entanglement in homogeneous
  fermionic chains},\ }\href {https://doi.org/10.1088/1751-8113/47/24/245301}
  {\bibfield  {journal} {\bibinfo  {journal} {Journal of Physics A:
  Mathematical and Theoretical}\ }\textbf {\bibinfo {volume} {47}},\ \bibinfo
  {pages} {245301} (\bibinfo {year} {2014})}\BibitemShut {NoStop}%
\bibitem [{\citenamefont {Murciano}\ \emph {et~al.}(2020)\citenamefont
  {Murciano}, \citenamefont {Ruggiero},\ and\ \citenamefont
  {Calabrese}}]{murciano2020symmetry}%
  \BibitemOpen
  \bibfield  {author} {\bibinfo {author} {\bibfnamefont {S.}~\bibnamefont
  {Murciano}}, \bibinfo {author} {\bibfnamefont {P.}~\bibnamefont {Ruggiero}},\
  and\ \bibinfo {author} {\bibfnamefont {P.}~\bibnamefont {Calabrese}},\
  }\bibfield  {title} {\bibinfo {title} {Symmetry resolved entanglement in
  two-dimensional systems via dimensional reduction},\ }\href
  {https://doi.org/10.1088/1742-5468/aba1e5} {\bibfield  {journal} {\bibinfo
  {journal} {Journal of Statistical Mechanics: Theory and Experiment}\ }\textbf
  {\bibinfo {volume} {2020}},\ \bibinfo {pages} {083102} (\bibinfo {year}
  {2020})}\BibitemShut {NoStop}%
\bibitem [{\citenamefont {Lundengard}(2017)}]{vandermonde}%
  \BibitemOpen
  \bibfield  {author} {\bibinfo {author} {\bibfnamefont {K.}~\bibnamefont
  {Lundengard}},\ }\bibfield  {title} {\bibinfo {title} {Generalised
  vandermonde matrices and determinants in electromagnetic compatibility},\
  }\href {https://api.semanticscholar.org/CorpusID:125673836} {\bibfield
  {journal} {\bibinfo  {journal} {Mälardalen University Press Licentiate
  Theses}\ ,\ \bibinfo {pages} {23}} (\bibinfo {year} {2017})}\BibitemShut
  {NoStop}%
\end{thebibliography}%

\end{document}